\documentclass[12pt,twoside]{article}
\usepackage{correl}
\usepackage{bm}

\def\TheTitle{Calculating Feynman diagrams with matrix product states}
\def\TheHeading{Feynman diagrams with MPS}
\def\TheAuthor{Xavier Waintal}
\def\TheInstitute{PHELIQS}
\def\TheAddress{PHELIQS, Université Grenoble Alpes, CEA, Grenoble INP, IRIG, Grenoble 38000, France}
\def\TheChapter{5}

\newcommand{\mI}{\mathcal{I}}
\newcommand{\mJ}{\mathcal{J}}
\newcommand{\Tr}{\mathop{\text{Tr}}}

\newcommand{\nd}[1]{ \mathrm{{\textbf{#1}}} }
\newcommand{\hnd}[1]{ \hat{\mathrm{{\textbf{#1}}}} }
\newcommand{\hid}[1]{ \tilde{\mathrm{{\textbf{#1}}}} }

\def\be{\begin{equation}}
\def\ee{\end{equation}}

\begin{document}
\MakeTitle

\section{Introduction}

Feynman diagrams form an exact formal solution of quantum field theories and quantum many-body problems in terms of a (huge) sum of (multi-dimensional) integrals of product of free propagators (which are known). 
For a long time these diagrams were mostly used and studied analytically. Countless
articles are devoted to e.g. analyse their structure or find subsequences of diagrams that could be summed exactly and form an appropriate approximate solution of the many-body
problem in this or that regime.
The idea to teach a computer how to calculate these diagrams is of course very appealing.
A particular route in this direction is known as ``diagrammatic Monte-Carlo''
\index{diagrammatic Monte-Carlo} and can be traced down to e.g. \cite{prokofiev1996,houcke2008} for equilibrium and \cite{werner2009} in the out-of-equilibrium context. Diagrammatic Monte-Carlo combines two ideas: First, since calculating a Feynman diagram amounts to calculating a multi-dimensional integral, and since Monte-Carlo (e.g. Metropolis algorithm) is the primary method for that,
one aims at sampling these integrals through a Markov process. Second, since there are a very large number of Feynman diagrams (typically $O(n!)$ for a calculation at order $n$), it is not possible to calculate them all so one might as well use the Markov process for both integration \emph{and} summing the diagrams. In other words, in diagrammatic Monte-Carlo, a configuration is a diagram with its set of vertices (times and positions where an interaction event occurs) and the Markov process introduce a random walk between different diagrams and different sets of vertices. The original version of diagrammatic Monte-Carlo had some successes but suffered from three difficulties:
\begin{itemize}
\item Problem A: The number of diagrams grows too fast with $n$, making it difficult to obtain converged results beyond $n=6-7$.
\item Problem B: The integrands may oscillate (depending on the regime) and become untractable through Monte-Carlo. This is known as the "sign problem" in quantum Monte-Carlo.
\item Problem C: The expansion itself may not converge even if large values of $n$ can be obtained. This depends on the nature of the series (assymptotic, finite radius of convergence...).
\end{itemize}
We will discuss these three aspects, in turn, in details.
The present notes describe a set of works 
\cite{profumo2015,bertrand2019,bertrand2020,macek2020,nunezfernandez2022,jeannin2024,jeannin2025}
my collaborators and I did to address these problems in the context of
quantum nanoelectronics. There are no new results here, merely this is a high level view
of what we did, designed to be much less formal and accessible than the original research articles. Among the different authors, these notes owe much to my old partner Olivier Parcollet who has been my constant collaborator on this topic. Most of the actual work, as often, has been carried out by young researchers including (using time ordering as is fit for a diagrammatic paper) Elio Profumo, Corentin Bertrand, Marjan Macek, Philipp Dumitrescu, Matthieu Jeannin, Thomas Kloss and Yuriel Nunez Fernandez. 

The most urgent problem was Pb.A and it was also how we got started on the topic.
In \cite{profumo2015}, we found a way to group the $n!$ diagrams into a much smaller sets of $2^n$ determinants of $n\times n$ matrices, dramatically reducing the complexity. This was done in the context of the Keldysh formalism suitable to treat out-of-equilibrium problems. A similar reduction in the context of imaginary time diagrams was found two years later in \cite{rossi2017}. These developments made calculations possible up to $n\sim 15$ or so in the absence of sign problem (Pb. B). 

To calculate the resulting integrals (Pb. B) we took a somewhat circonvoluted path. We started with Monte-Carlo sampling as was done in diagrammatic Monte-Carlo \cite{profumo2015,bertrand2019}. It became quickly obvious that this was inefficient: the integrand was called typically $10^9$ times in a practical calculation (each call having a cost of up to $O(2^{15})$ floating operations for the largest $n=15$ that we could reach) but we did not take any advantage of the associated accumulated information. 
An initial idea was that an approximation of the integrand could be learned along the way 
(machine learning is the fashion). In turn, this approximation could be used to either speed-up the Markov process (decrease its correlation time through smarter proposed moves) or, as it turned out, obtain better convergence using low discrepany sequences (better known as quasi Monte-Carlo) \cite{macek2020}. We shall not follow this line of thought here because it was eventually supplanted by a much better technique. Indeed, we found that the integrand could be learned directly using tensor network techniques and that the resulting tensor network could be integrated exactly without any need for any Monte-Carlo \cite{nunezfernandez2022}.
Hence, the technique in its current form does not use diagrams anymore and does not use Monte-Carlo either. It is a sort of "Non-diagrammatic Non-Monte-Carlo" technique.

For the last problem C, we took two different routes. The first was to consider the resummation as an analytical continuation problem \cite{profumo2015,bertrand2020} which was effective in some situations but difficult to turn into an automatic technique (i.e. that works without human supervision). Here, I will focus on our latest approach that uses 
``cross extrapolation'' \cite{jeannin2024} and that is routed in the same mathematics as our tensor network learning technique.

In the remaining of these notes, I will walk the reader through what I just described i.e.
describing the problem, then our technique to try and solve problem A, then problem B and ultimately problem C. I will end by showing some actual data in the context of the out-of-equilibrium Anderson model (SIAM), the main model for which we have had results at the time of this writing \cite{jeannin2025}. Our main success there was to be able to compute the differential conductance versus bias and gate voltages ``exactly'' including its most proeminent features, the ``Kondo ridge'' and the ``Coulomb diamonds''. By "exactly" here, we means that the technique is controlled: it has error bars that (i) are known and (ii) can be systematically improved by increasing the computing time (in contrast to e.g. mean field techniques).  

\section{Problem formulation}

\subsection{General model}
\label{sec:general_model}
The type of models we want to study correspond to a finite interacting 
quantum nanoelectronic system connected to infinite (metallic) electrodes, following the approach of Ref.~\cite{meir1992}. The Hamiltonian
consists of a quadratic term and an electron-electron interaction term,
\begin{equation}
\hnd{H}(t)=\hnd{H}_0(t)+U \hnd{H}_{\rm int}(t)
\label{eq:H_general}
\end{equation}
where the parameter $U$ controls the magnitude of the interaction.
The non-interacting Hamiltonian takes the following form,
\be
\hnd{H}_0(t)=\sum_{i,j} H^0_{ij}(t)\hnd{c}^\dagger_{i}\hnd{c}_{j}
\ee
where $\hnd{c}^{\dagger}_{i}$ ($\hnd{c}_{j}$) are the usual fermionic creation (annihilation)
operators of a one-particle state on the site $i$. 
\begin{equation}
\{ \hnd{c}^{\dagger}_{i} , \hnd{c}_{j}  \} = \delta_{ij}
\end{equation} 
The site index $i$ is general and can include different kinds of degrees of freedom: space, spin, orbitals.
A crucial aspect is that the number of ``sites'' is {\it infinite} so that the non-interacting system has a well-defined density of states (as opposed to a sum of delta functions for a finite system) while interactions only take place in a finite region. 
This will ensure the infrared convergence of the various terms of the perturbation expansion while using a discretized system provides ultraviolet convergence. So in contrast to quantum field theory where some diagrams may diverge, here each diagram taken separately will be finite. The interaction Hamiltonian takes the form
\be
\hnd{H}_{\rm int}(t)=\sum_{ijkl} V_{ijkl}(t)\hnd{c}^\dagger_{i}\hnd{c}^\dagger_{j}\hnd{c}_{k}\hnd{c}_l
\ee
In contrast to the non-interacting part, it is confined to a {\it finite} region. We also supposed that the interaction vanishes for negative time and is slowly or abruptly switched on at $t=0$. 

\subsection{Summary of the overall approach}
\label{sec:overal_approach}

\begin{figure}
	\centerline{\includegraphics[scale=0.9]{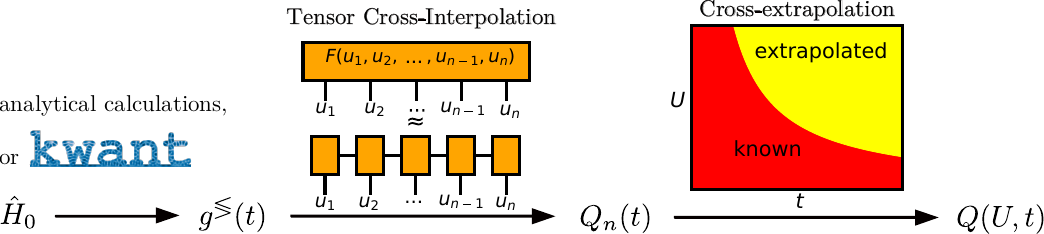}}
	\caption{Schematic of the overall approach. 
Starting from non interacting Green's functions, a decomposition of
the multidimensional integral is obtained using Tensor Cross-Interpolation
to compute the perturbative expansion, from which the physical observables
$Q(U,t)$ are reconstructed as a function of interaction $U$ and time $t$ using
cross-extrapolation. Adapted from \cite{jeannin2025}}
	\label{fig:method}
\end{figure}

Our goal is to perform a systematic expansion of some observable in power of $U$.
For instance, if we are interested in the occupation $Q(U,t)$ of a given site $i_0$
at time $t$, we will write its expansion as,
\begin{equation}
Q(U,t) \equiv \langle\hnd{c}^\dagger_{i_0}\hnd{c}_{i_0} \rangle = 
\sum_{n=0}^{+\infty} Q_n(t) U^n.
\label{eq:series}
\end{equation}
Our goal will be to first compute the coefficients $Q_n(t)$ for as large a $n$ as possible
and then sum up the series. A schematic of the method is shown in Fig.\ref{fig:method}.
It consists of three main steps: {(\it i)} solve the non-interacting model to obtain the corresponding non-interacting Green's functions (such as the lesser $g_{ij}^<(t)$ or
upper $g_{ij}^>(t)$ green functions that will be introduced below).
{(\it ii)} Calculates the coefficients $Q_n(t)$ with high precision 
and {(\it iii)} resumming this perturbative expansion.

The first step is carried out either analytically in simple models or numerically for more complex geometries using e.g. the Kwant \cite{groth2014} or Tkwant \cite{kloss2021} software. Note that the non-interacting Green's function may already be out-of-equilibrium in presence of a bias voltage $V_b$.

The second step is to perform the actual expansion in power of $U$. We shall see that this
expansion gives an $n$-dimensional integral that contains $2^n$ terms constructed out of products of the non-interacting Green's functions \cite{profumo2015,bertrand2020}. We compute this integral using the Tensor Cross-Interpolation (TCI) algorithm which vastly outperforms previous quantum Monte-Carlo and quasi Monte-Carlo approaches for this problem
\cite{nunezfernandez2022}. A large fraction of these notes will be devoted to an introduction to TCI which has applications far beyond the present problem. See \cite{nunezfernandez2025}  for an in-depth description of TCI. An associated open source library may also be found at https://tensor4all.org. With this we will be able to calculate all Feynman diagrams 
up to $n=N$ with typically $N=20-25$ for our example problem.

The last step of the method consist in reconstructing the function $Q(U,t)$, if possible for both short and long times $t$ and interaction strength $U$ from the knowledge of only $N$
coefficients $Q_n(t)$. The analytical structure of $Q(U,t)$ is quite interesting:
At any fine time $t$, it has an infinite radius of convergence \cite{bertrand2019} since
\be
Q_n \sim \frac{t^n}{n!} \rightarrow  Q_nU^n \sim \left(\frac{e U t}{n}\right)^n.
\ee 
which implies that $N \sim O(Ut)$ terms are required to converge the sum using the finite sum
\begin{equation}
Q(U,t) \approx \sum_{n=0}^N Q_n(t) U^n.
\label{eq:series_approx}
\end{equation}
On the other hand, in the steady state $t\rightarrow\infty$, it has a finite radius of convergence $R$ and $Q_n\sim 1/R^n$ and the above naive summation fails for $U>R$.
In other words, the two limits $t\rightarrow\infty$ and $N\rightarrow\infty$ do not commute.
Here, we will focus on one approach to perform the reconstruction,
the {\it cross-extrapolation} \cite{jeannin2024}. It is based on two ideas,
\begin{itemize}
\item First, using the naive sum we can obtain $Q(U,t)$ in two different regimes
\begin{itemize}
\item Arbitrary times and small interactions $U<R$.
\item Large interactions but small times ($U\sim N/t$).
\end{itemize}
To calculate $Q(U,t)$ at both large $t$ \emph{and} $U$, it is therefore tempting to perform a \emph{double} extrapolation from these two limits simultaneously.
\item Second, we will use the fact that $Q(U,t)$ \emph{almost} factorizes. i.e. seen as a matrix where $U$ are the rows and $t$ the columns, it is of low rank.
\end{itemize}
As we shall see, cross-extrapolation is a general idea that could be used in many other situations.

\subsection{The SIAM model}
\begin{figure}
	\centering
	\includegraphics[width=0.45\textwidth]{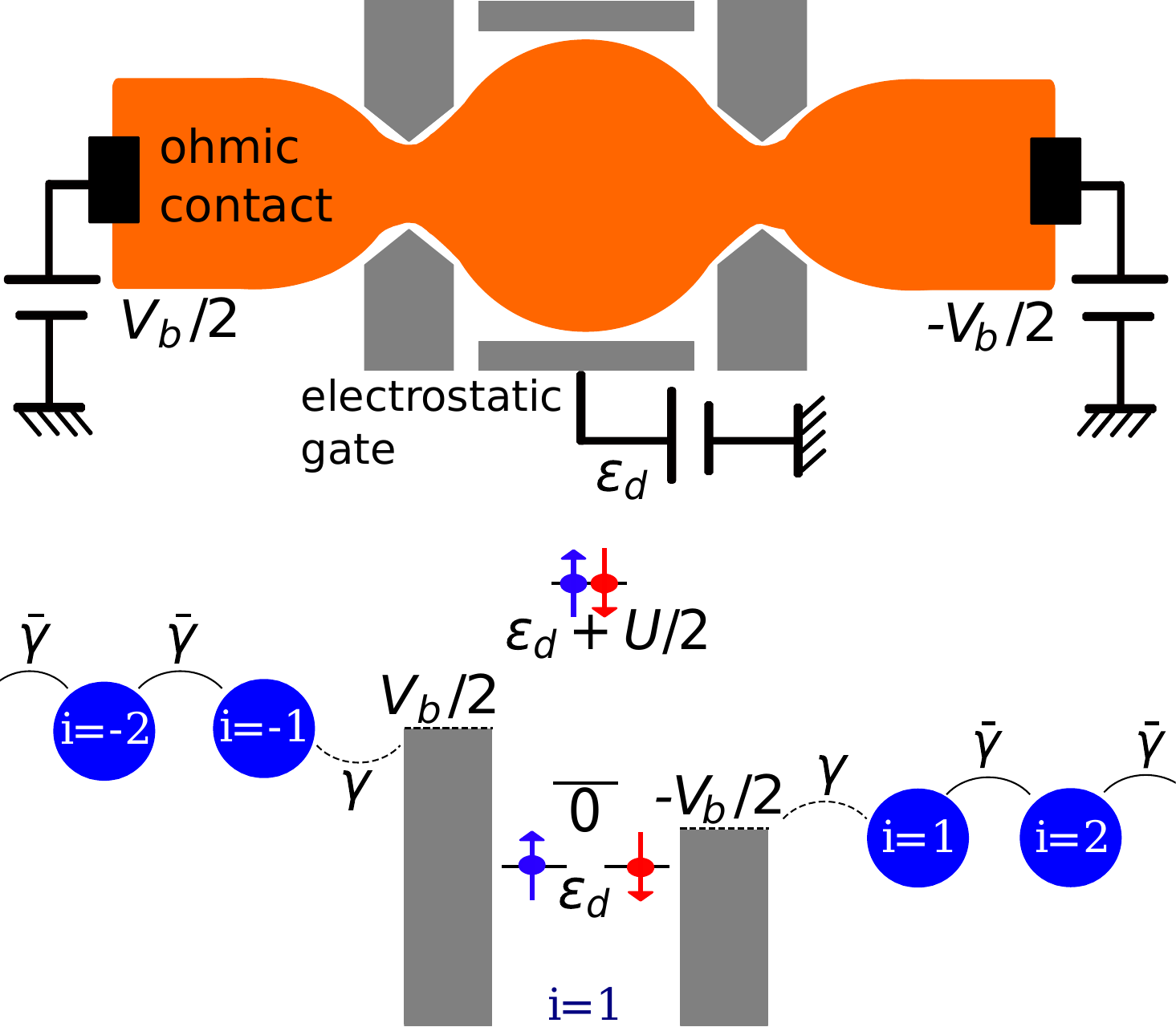}
\caption{Top: electrical diagram of the quantum dot coupled to two electrodes.
Bottom: schematic of the SIAM. Adapted from \cite{jeannin2025}.
\label{fig:siam} }
\end{figure} 
Although the approach is in principle general, in practice we will demonstrate it for a concrete model, the out-of-equilibrium SIAM. The model corresponds to a quantum dot, typically constructed by confining electrons with electrostatic gates inside a semiconductor
or a semiconductor heterostructure. The quantum dot is modeled by a single interacting level 
(site $i=0$) weakly connected to two semi-infinite one dimensional electrodes ($i<0$ and $i>0$, respectively) at chemical potential $\mu_{l/r} = \pm V_b/2$, as
depicted in Fig.~\ref{fig:siam}. Its Hamiltonian reads

\begin{equation}
\hnd{H} =  
\sum_{i=-\infty}^{+\infty} 
\sum_{\sigma \in \{\uparrow, \downarrow \} } 
\left(	\gamma_i \hnd{c}^\dagger_{i,\sigma }\hnd{c}_{i+1,\sigma } +\text{h.c.}\right)
+\epsilon_d(\hnd{n}_\uparrow+\hnd{n}_\downarrow) + U \theta(t) \hnd{n}_\uparrow\hnd{n}_\downarrow
\end{equation}
where the former sites $i$ are now extended to include the spin $\sigma = \uparrow, \downarrow$ degree of freedom. $\epsilon_d$ is the on-site energy on the dot controlled by an electrostatic gate and $\hnd{n}_{\sigma}= \hnd{c}^\dagger_{0,\sigma}\hnd{c}_{0,\sigma}$. 
We work directly in the thermodynamic limit with an infinite number of bath sites. 
$\theta(t)$ is the Heaviside function.
The hopping between two neighbouring sites is denoted by $\gamma_i=\bar{\gamma}$,
except for the coupling to the dot $\gamma_0=\gamma_{-1}=\gamma$. 
Energies are expressed in unit of the tunneling rates to the dot $\Gamma = 2\gamma^2/\bar{\gamma}$. For concretness, all calculations are performed at zero temperature. 

Our goal will be to calculate the charge $Q$ on the dot and the current $I$ flowing from the dot to the right electrode after an interaction quench at $t=0$.
\begin{align}
	Q(U,t) &= \sum_{\sigma \in \{\uparrow, \downarrow \} }\langle \hnd{c}_{0,\sigma}^\dagger 
	\hnd{c}_{0,\sigma} \rangle \\	
	I(U,t) &=  -i \gamma  \sum_{\sigma \in \{\uparrow, \downarrow \} } \left[ \langle \hnd{c}_{1,\sigma}^\dagger \hnd{c}_{0,\sigma} \rangle - \langle \hnd{c}_{0,\sigma}^\dagger \hnd{c}_{1,\sigma} \rangle\right] \label{eq:current_def}.
\end{align}

\section{Interaction expansion (problem A)}
 
 This section explains the first step of the approach: how to reduce the calculation of
 $Q_n(t)$ to a multidimensional integral in terms of known objects.
 
\subsection{A short reminder of Keldysh formalism}
Our starting point for this work is a formal expansion of the out-of-equilibrium (Keldysh) Green's function in powers of electron-electron interactions. This is a standard step\cite{rammer1986} which we briefly sketch to introduce our notations. We would also like to demistify Keldysh formalism which is much simpler than often thought to be. In our opinion it is actually conceptually simpler than imaginary time or zero temperature formalism at the price of the actual calculations being a bit more cumbersome. Since these calculations will be performed by the computer, this is not necessarily an issue.

One starts by the standard step of "integrating out'' the (supposingly known) dynamics of
$\hnd{H}_0$, i.e. work in the interaction representation.
Using the interaction representation, one defines $\hnd{c}_i(t)=\hnd{U}_0(0,t)\hnd{c}_i \hnd{U}_0(t,0)$ where $\hnd{U}_0(t',t)$ is the evolution operator from $t$ to $t'$ associated with $\hnd{H}_0$. For a time independent $\hnd{H}_0$, 
it is simply $\hnd{U}_0(t',t) = e^{-i\hnd{H}_0(t'-t)}$. In this representation,
one defines $\hid{H}_{\rm int}(\bar u)$, which is equal to $\hnd{H}_{\rm int}(u)$ with the operators $\hnd{c}_i,\hnd{c}^\dagger_j$ replaced by $\hnd{c}_i(\bar u),\hnd{c}^\dagger_j(\bar u)$. Starting from a non-interacting density matrix $\rho_0$ at $t=0$,
the average of an observable $\hnd{O}$ at time $t$ is given by
\be
\langle \hnd{O} \rangle = \Tr  \left[ T e^{ +i \int d\bar u\ U \hid{H}_{\rm int}(\bar u)} 
\ \hid{O}\ T e^{ -i \int d\bar u\ U \hid{H}_{\rm int}(\bar u)} \rho_0 \right]
\label{eq:keldyshexplained}
\ee
where $T$ is the usual time-ordering operator (now necessary because in the interaction representation the interacting Hamiltonian \emph{is} time dependent even if the original Hamiltonian isn't). The above equation is very natural; for instance if 
$\rho_0 = |\Psi_0\rangle\langle \Psi_0|$ then it corresponds to 
$\langle \hnd{O} \rangle = \langle \Psi(t) | \hnd{O}|\Psi(t) \rangle$ with 
$|\Psi(t) \rangle = T e^{ -i \int d\bar u\ U \hid{H}_{\rm int}(\bar u)} |\Psi_0 \rangle$
the state of the system at time $t$. Looking at Eq.(\ref{eq:keldyshexplained}), we see that when we will perform the expansion in power of $U$, there will be two kind of terms: the one coming before the operator $\hnd{O}$ and the one coming after it. To remember if a terms originates from one or the other (before of after), one introduces an index $a=0$ (before) or $a=1$ (after). This is the "Keldysh index" and that's all there is to it: Keldysh formalism is merely a technique to book-keep the position of the different terms in the expansion.

A bit more formally, one defines the contour ordering for pairs $\bar t=(t,a)$: $(t,0)<(t',1)$ for all $t,t'$, $(t,0)<(t',0)$ if $t<t'$ and $(t,1)<(t',1)$ if $t>t'$. The contour ordering operator $T_c$ acts on products of fermionic operators $A,B,C\dots$ labeled by various ``contour times'' $\bar t_A=(t_A,a_A),\bar t_B,\bar t_C\dots$ and reorder them according to the contour ordering: $T_c(A(\bar t_A)B(\bar t_B)=AB$ if  $\bar t_A>\bar t_B$ and $T_c(A(\bar t_A)B(\bar t_B)=-BA$ if  $\bar t_A<\bar t_B$. The non-interacting contour Green's function is defined as
\be
g^c_{ij}(\bar t,\bar t')= -i \langle T_c \hnd{c}_i(\bar t) \hnd{c}^\dagger_j(\bar t') \rangle
\ee
where $\hnd{c}_i(\bar t)$ is just $\hnd{c}_i(t)$, the Keldysh index serving only to define the position of the operator after contour ordering. The contour Green's function has a matrix structure in $a,a'$ which reads
\be
g^c_{ij}(t,t')=
\left(
\begin{array}{cc}
g^T_{ij}(t,t') & g^<_{ij}(t,t') \\
g^>_{ij}(t,t')   &  g^{\bar T}_{ij}(t,t')
\end{array}
\right)
\ee
where $g^T_{ij}(t,t')$, $g^<_{ij}(t,t')$, $g^>_{ij}(t,t')$ and $g^{\bar T}_{ij}(t,t')$ are respectively the
time ordered, lesser, greater and anti-time ordered Green's functions. 
These non-interacting Green's functions will form the actual input of our approach,
we will briefly discuss how they are obtained in the next section.
Finally, one defines the full Green's function 
$G^c_{ij}(\bar t,\bar t')$ with definitions identical to the above except that $\hnd{U}_0$ is replaced by $\hnd{U}$, the
evolution operator associated to the full Hamiltonian $\hnd{H}$. 
The fundamental expression for $G^c_{ij}(\bar t,\bar t')$ reads
\be
G^c_{ij}(\bar t,\bar t')= -i\langle T_c e^{ -i \int d\bar u\ U \hid{H}_{\rm int}(\bar u)} \hnd{c}_i(\bar t) \hnd{c}^\dagger_j(\bar t') \rangle
\ee
where the integral over $\bar u$ is taken along the Keldysh contour, i.e. increasing $u$ for $a=0$ and decreasing for $a=1$. We are almost ready to perform the expansion. 

\subsection{Non-interacting Green's functions}
The dynamics of the non-interacting problem is, in principle, "trivial" in the sense that one simply needs to solve the one-body problem and "fill up" the states up to the Fermi energy. In practice it may not be entirely straightfoward but there are well known and mature techniques to calculate both the stationary \cite{waintal2024} and time-dependent properties \cite{gaury2014}. There are also associated open source softwares such as Kwant \cite{groth2014} for the stationary problem and TKwant \cite{kloss2021} for the time dependent one. The latter explicitly supports that calculation of $g^T_{ij}(t,t')$, $g^<_{ij}(t,t')$, $g^>_{ij}(t,t')$ and $g^{\bar T}_{ij}(t,t')$.

One approach is to relate the non-interacting Green's functions 
to the (Scattering) wave functions in the system \cite{gaury2014},
\begin{align}
&g_{ij}^<(t,t') =i\sum_\alpha \int \frac{dE}{2\pi}\ f_\alpha(E)  \Psi_{\alpha E}(t,i) \Psi^*_{\alpha E}(t',j)
\label{eq:psi-less}
\end{align}
Here, $\alpha$ labels the various propagating channels of the leads, $\Psi_{\alpha E}(t,i)$ the scattering state at energy $E$ (in the electrode) and $f_\alpha(E)$ the corresponding Fermi distribution function.
The greater Green's function $g_{ij}^>(t,t')$ is obtained with an identical expression with the Fermi functions 
$f(E)$ replaced by $f(E)-1$. 
The actual calculations performed in this article were restricted to a stationary non-interacting system, where
the above expression further simplifies into
\begin{align}
&g_{ij}^<(t-t') =i\sum_\alpha \int \frac{dE}{2\pi}\ f_\alpha(E)  \Psi_{\alpha E}(i) \Psi^*_{\alpha E}(j)e^{-iE(t-t')}
\label{eq:psi-less-st}
\end{align}
Here again, the stationary scattering wave functions $\Psi_{\alpha E}(i)$ are standard objects \cite{waintal2024}. They are the eigenvectors of $\hnd{H}_0$ with one caveat: they need to be classified into states incoming from the left and states incoming from the right. These objects are in fact direct outputs of the Kwant software \cite{groth2014}. Once the lesser and greater Green's functions are known, one completes the $2\times 2$ Keldysh matrix with the standard relations
\begin{align}
&g^T_{ij}(t,t')= \theta(t-t')g^>_{ij}(t,t') +
\theta(t'-t)g^<_{ij}(t,t')\\
&g^{\bar T}_{ij}(t,t')= \theta(t'-t)g^>_{ij}(t,t') +
\theta(t-t')g^<_{ij}(t,t')
\end{align}

\subsection{The expansion itself}
\begin{figure}[h]
    \centering
    \includegraphics[width=7cm]{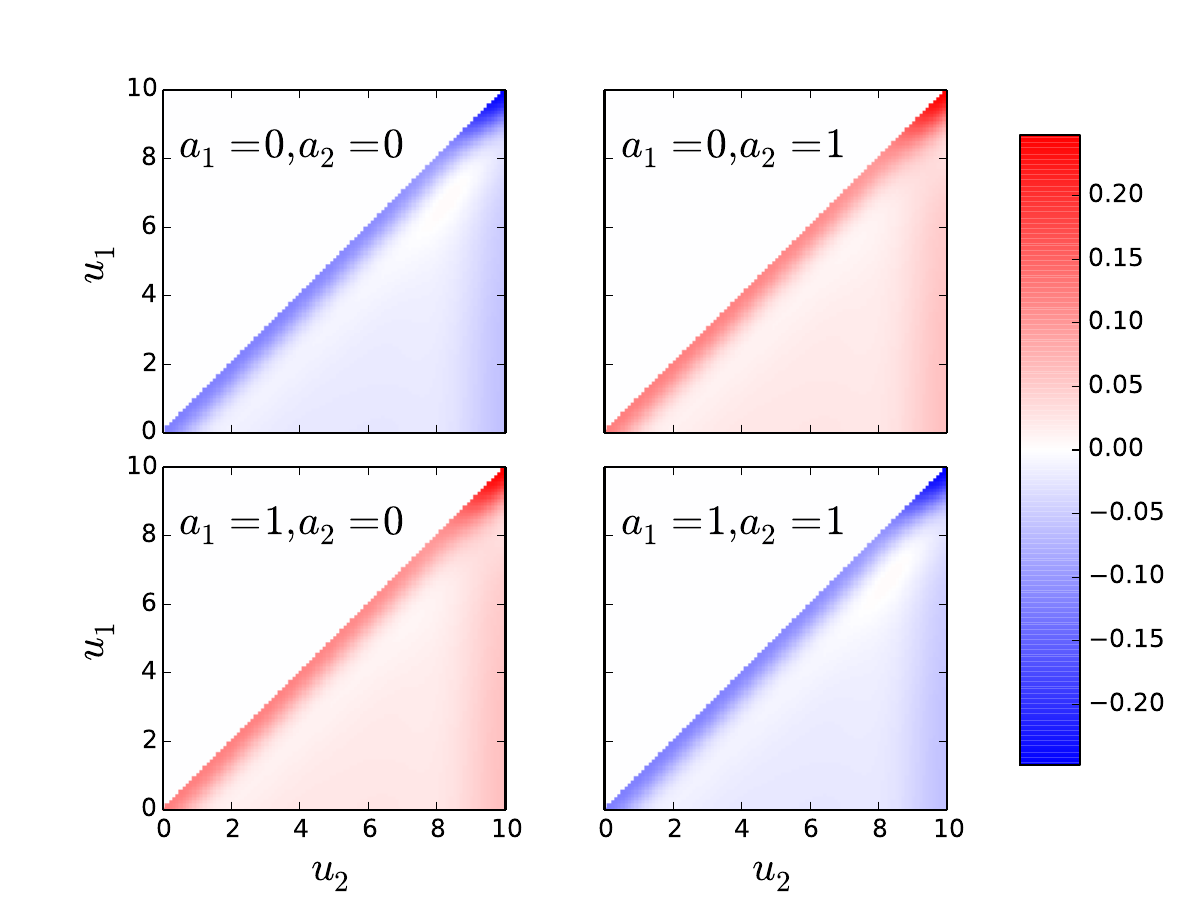}
     \includegraphics[width=7cm]{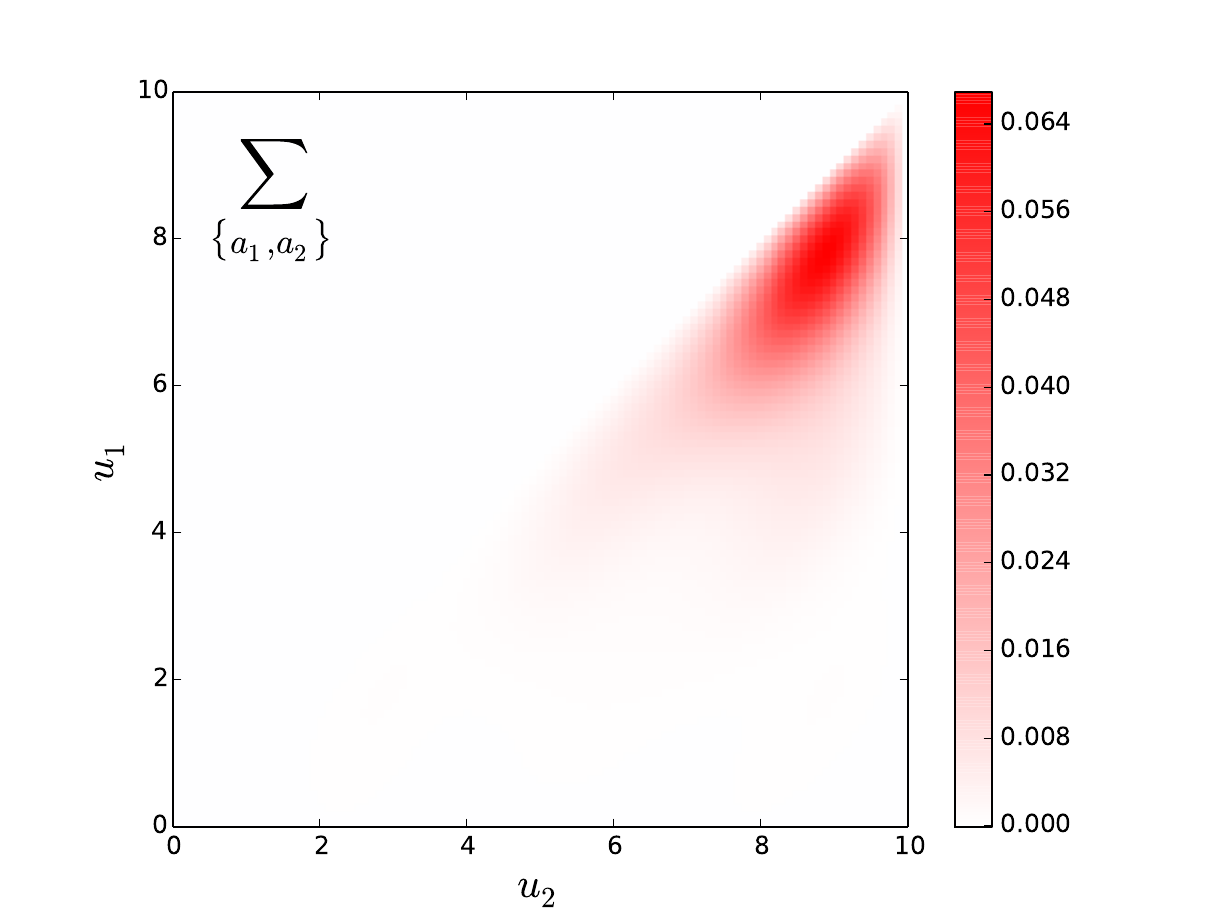}
    \caption{\label{fig:M} Left: Colorplot of the integrand  of $Q_2$ as a function of the two times $u_1$ and $u_2$ for SIAM with $\mu_L=\mu_R=0$, $\epsilon_d=0$, $T=0$ and $t=10$. 
    The four panels correspond to the 4 possible values of the two Keldysh indices $a_1$ and $a_2$. The explicit form of the integrand is $f(u_1,u_2,a_1,a_2)= -\Im m (-1)^{\sum_i a_i}  \det \nd{M}_{2}(u_1,u_2,a_1,a_2)$. Right: Same parameters as on the left but the integrand has now been summed over Keldysh indices. The colorplot represents $f(u_1,u_2)= i \sum_{a_1,a_2} (-1)^{\sum_i a_i}  \det \nd{M}_{2}(u_1,u_2,a_1,a_2)$
 ($f$ is real). Note that the integrand is now real, positive and concentrated around $u_1 = u_2 = t$. Adapted from \cite{profumo2015}
    }
\end{figure}

We continue to follow the literature and performs the expansion in powers of $U$. We obtain:
\be
G^c_{ij}(\bar t,\bar t')= -i\sum_{n=0}^{+\infty} \frac{(-i)^n}{n!} U^n \sum_{\{a_i\}} (-1)^{\sum_i a_i}
\int du_1du_2\dots du_n \langle T_c \hid{H}_{\rm int}(\bar u_1) \hid{H}_{\rm int}(\bar u_2)\dots \hid{H}_{\rm int}(\bar u_n) \hnd{c}_i(\bar t) \hnd{c}^\dagger_j(\bar t') \rangle
\ee
and we are left to calculate the non-interacting average of a (possibly large) product of
creation and destruction operators. This is done, as usual, using Wick theorem. Wick theorem states that an average over a non-interacting density matrix of a product of fermionic operators
takes the form,
\be
\langle \hnd{c}_1^\dagger \hnd{c}_1 \hnd{c}_2^\dagger \hnd{c}_2 ...
\hnd{c}_M^\dagger \hnd{c}_M \rangle
=
\sum_P (-1)^{|P|}
\langle \hnd{c}_1^\dagger \hnd{c}_{P(1)} \rangle
\langle \hnd{c}_2^\dagger \hnd{c}_{P(2)} \rangle ...
\langle \hnd{c}_M^\dagger \hnd{c}_{P(M)} \rangle
\label{eq:wickdet1}
\ee
where the sum runs over all the permutation $P$ of $M$ elements, 
$|P|=\pm 1$ is its signature
and $P(a)$ the image of $a$ through the corresponding permutation. The numbers
$1,2...M$ are just shorthands for all the parameters of the fermionic operators:
sites $i$ (including spin), times $u$ and Keldysh index $a$. 
 
This is where the numerical route starts to differ from the analytical one. 
Usually, one would identify each permutation with a Feynman diagram and starts do derive the corresponding Feynman rules. But as already discussed, the problem is that we would have
far too many diagrams (there are $M!$ permutations in the above expression). For numerical purposes, we will make a simple remark: that the right hand side of Eq.(\ref{eq:wickdet1})
is actually the definition of the determinant of a $M\times M$ matrix:
\be
\langle \hnd{c}_1^\dagger \hnd{c}_1 \hnd{c}_2^\dagger \hnd{c}_2 ...
\hnd{c}_M^\dagger \hnd{c}_M \rangle
=
\text{det} \langle \hnd{c}_a^\dagger \hnd{c}_{b} \rangle.
\ee
Since the calculation of the determinant of a $M\times M$ matrix takes only 
$M^3\ll M!$ operations, this sounds very appealing. Now putting things together, we arrive
at a formula that is conceptually very simple: the left hand side is what we want and the right hand side is a set of multi-dimensional integrals of matrices whose entries are known.
The problem is therefore "reduced to quadrature",
\be
G^c_{ij}(\bar t,\bar t')= \sum_{n=0}^{+\infty} \frac{i^n}{n!} U^n \sum_{\{a_i\}} (-1)^{\sum_i a_i}
\int du_1du_2\dots du_n
\sum_{i_1j_1k_1l_1}V_{i_1j_1k_1l_1}(u_1)\dots \sum_{i_nj_nk_nl_n}  V_{i_nj_nk_nl_n}(u_n)
\det  \nd{M}_n
\label{eq:basic}
\ee
where the $(2n+1) \times (2n+1)$ matrix $\nd{M}_n$ is given by
\be
\label{eq:M}
\nd{M}_{n} =\left(
\begin{array}{lllll} 
g^<_{k_1i_1}(\bar u_1,\bar u_1) & g^>_{k_1j_1}(\bar u_1,\bar u_1)  & g^c_{k_1i_2}(\bar u_1,\bar u_2) &...&g^c_{k_1 j}(\bar u_1,\bar t')  \\
g^<_{l_1i_1}(\bar u_1,\bar u_1) & g^<_{l_1j_1}(\bar u_1,\bar u_1)  & g^c_{l_1i_2}(\bar u_1,\bar u_2) &...&g^c_{l_1 j}(\bar u_1,\bar t') \\
g^c_{k_2i_1}(\bar u_2,\bar u_1) & g^c_{k_2j_1}(\bar u_2,\bar u_1)  & g^<_{k_2i_2}(\bar u_2,\bar u_2) &...&g^c_{k_2 j}(\bar u_2,\bar t')  \\
...&...&...&...&...\\
g^c_{k_ni_1}(\bar u_n,\bar u_1) & g^c_{k_nj_1}(\bar u_n,\bar u_1)  & g^c_{k_ni_2}(\bar u_n,\bar u_2) &...&g^c_{k_n j}(\bar u_n,\bar t')  \\
g^c_{l_ni_1}(\bar u_n,\bar u_1) & g^c_{l_nj_1}(\bar u_n,\bar u_1)  & g^c_{l_ni_2}(\bar u_n,\bar u_2) &...&g^c_{l_n j}(\bar u_n,\bar t')  \\
g^c_{i i_1}(\bar t,\bar u_1) & g^c_{i j_1}(\bar t,\bar u_1)  & g^c_{i i_2}(\bar t,\bar u_2) &...&g^c_{i j}(\bar t,\bar t')  \\
\end{array}
\right)
\ee 
and the zeroth order term is $g^c_{ij}(\bar t,\bar t')$. [Note that there was a typo in the above expression in \cite{profumo2015}, see \cite{bertrand2019} for the correction].
To calculate an actual observable, we just recognize that it corresponds to the
lesser Green's function at equal times, i.e.
\be
\label{eq:O}
\langle \hnd{U}(0,t) \hnd{c}^\dagger_i\hnd{c}_j \hnd{U}(t,0) \rangle=-i G^<_{ji}(t,t)
\ee

We are just left with a simple problem: to obtain $Q_n(t)$ we must integrate over
$n$ times and sum over $n$ Keldysh indices (in general we also need to sum over the $n$
different vertices $V_{ijkl}$ but in the case of SIAM, this sum reduces to a single term). There is a caveat however. To see it, let us look at the integrand of $Q_2$ in the
$4$-sectors defined by the two Keldysh indices, as shown in the left panel of Fig.\ref{fig:M}. We immediatly see that the corresponding integrals are not going to behave well: the integrand does not seem to decay when the time gets away from the time where the measurement is made
(reminder: we switch on the interaction at $t=0$ and measure, here, at $t=10$). This is to be expected and is why the determinant form of the Wick theorem was not super successful before: the expansion includes all Feynman diagrams, \emph{including} the disconnected diagrams. It is well known that these disconnected diagrams do not contribute to the final result. Yet, in this expansion they are present and a priori only cancel at the end of the calculation after integration and summation over Keldysh indices.

The magic occurs when one performs the summation over the Keldysh indices, as shown in
the right panel of Fig.\ref{fig:M}: we find that the cancelation of the disconnected diagrams occurs \emph{before} the integration over times. The proof is a bit technical and is shown in the appendix of \cite{profumo2015}. The consequence of this property is that, one is left with a well behaved integral to calculate. The cost of each call to the integrand is $O(2^n)$ which is still exponential but very mild compared to the initial $n!$ cost we started with. To calculate these integrals we will use Tensor Cross Interpolation which we now explain.

\section{Tensor Cross interpolation for integration (problem B)}

 We will now discuss a very important algorithm - Tensor Cross Interpolation (TCI) - that has a very special place in the zoo of tensor network algorithms.
 \index{Tensor Cross Interpolation}
This algorithm takes as an input a "virtual" tensor 
$F_{\sigma_1,\sigma_2...\sigma_{N} }$ where each index $\sigma_i$ takes $d_i$ different values (for simplicity we assume that $d_i\equiv d$ is constant). It returns as an output a Matrix Product State (MPS) that approximates $F_{\vec{\sigma}}$ in the best possible way.
For an introduction to MPS, see \cite{schollwock2011}. For the purpose of this article, the MPS is merely the following expression:
\begin{equation}
F_{\sigma_1,\sigma_2...\sigma_{N} } \approx \sum_{\{\alpha_i\}} 
M^1_{\alpha_1}(\sigma_1)
M^2_{\alpha_1\alpha_2}(\sigma_2)...
M^{N}_{\alpha_{N-1}}(\sigma_N)
\end{equation}
where the matrices $M_a(\sigma)$ have a maximum size $\chi$, known as the bond dimension.
$F_{\vec{\sigma}}$ is virtual in the sense that the input of the algorithm is \emph{not} the actual tensor (which would be an exponentially large object with $d^N$ elements). Rather it is a function that takes 
$\vec{\sigma}= (\sigma_1,\sigma_2...\sigma_{N})$ as an input and
returns the corresponding value $F_{\vec{\sigma}}$. TCI is very different from many other tensor network algorithms (e.g. DMRG \cite{schollwock2011}): here $F_{\vec{\sigma}}$ is actually known by the user, what is not known is its MPS representation. Once one has this MPS, one can use it for calculating e.g. integrals or plenty of other applications that we shall not discuss, see \cite{nunezfernandez2025}. TCI is really a ``gateway'' that allows one to take a problem that is \emph{not} formulated in terms of tensor network and transform it into this framework. 

Another peculiarity of TCI is that it is a \emph{learning} algorithm akin to what is done in machine learning. More precisely it is an active learning algorithm since TCI decides on the data $(\vec{\sigma},F_{\vec{\sigma}})$ that will be requested. As in machine learning, only a very tiny fraction of the possible configurations $\vec{\sigma}$ will be explored, and as in machine learning the fact that the resulting model interpolates correctly between the configurations can be spectacular. On the other hand there are strong differences with deep neural networks: the optimization has nothing to do with gradient descent (and is way more effective) and the resulting function much more structured (for instance we can easily calculate e.g. integrals, we can't do that with a neural network). The cost for these added features is a more restrictive set of applications: TCI is only effective for problems where the level of ``entanglement'' is limited (small $\chi$).

The presentation below is mostly based on section III of \cite{nunezfernandez2022} with a few more advanced aspects borrowed from \cite{nunezfernandez2025} to which we also refer for the references to the original literature. 
The readers can also have a look at the tensor4all open source library that implements these algorithms https://tensor4all.org.

\subsection{Compressing low rank matrices}

\begin{figure}[htb]
\includegraphics[width=1.0\textwidth]{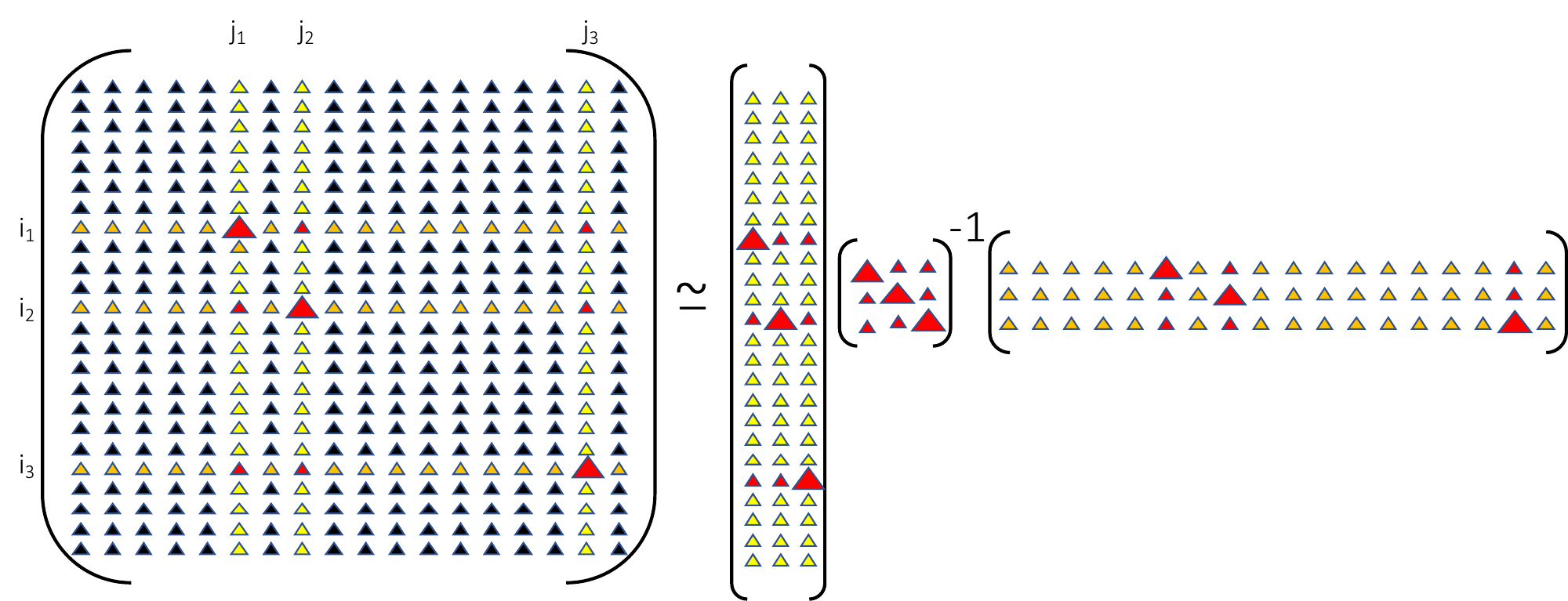}
\caption{\label{fig:ci}
   Illustration of the cross interpolation (CI) of a matrix.
   The large red triangles indicate real pivots and the smaller red
   triangles indicate automatically generated pivots. The right-hand side only contains small subparts of the matrix. Adapted from \cite{nunezfernandez2022}}
\end{figure}

Before we can get into TCI, we need a matrix factorization formula for low rank 
(or approximately low rank) matrices that is based Gaussian elimination and the concept of
Schur complement \cite{golub1996}. This formula (the ``cross interpolation'') will be almost as good as the Singular Value Decomposition (SVD, which is optimum) but with a key advantage: it can be performed without the need to access the full matrix $A$: only a set of $\chi$ rows and columns will be needed.

\subsubsection{Revisiting Gaussian elimination}

We consider an arbitrary matrix $A$ that we put in a $2\times 2$ block form,
\begin{equation}
A =
\begin{pmatrix}
A_{11} & A_{12} \\
A_{21} & A_{22}
\end{pmatrix}
\end{equation} 
Following the strategy of Gaussian elimination, we can put this matrix in triangular form as (provided the $A_{11}$ block is invertible), 
\begin{eqnarray}
\begin{pmatrix}
1 & 0 \\
- A_{21} A_{11}^{-1} & 1
\end{pmatrix}
\begin{pmatrix}
A_{11} & A_{12} \\
A_{21} & A_{22}
\end{pmatrix}
 \nonumber \\
=
\begin{pmatrix}
A_{11} & A_{12} \\
0 & A_{22} - A_{21}A_{11}^{-1} A_{12}
\end{pmatrix}.
\end{eqnarray}
We can proceed on the columns to eliminate the $A_{12}$ block and finally obtain,
\begin{eqnarray}
\begin{pmatrix}
1 & 0 \\
- A_{21} A_{11}^{-1} & 1
\end{pmatrix}
\begin{pmatrix}
A_{11} & A_{12} \\
A_{21} & A_{22}
\end{pmatrix}
\begin{pmatrix}
1 & -A_{11}^{-1} A_{12} \\
0 & 1
\end{pmatrix} \nonumber \\
=
\begin{pmatrix}
A_{11} & 0 \\
0 & A_{22} - A_{21}A_{11}^{-1} A_{12}
\end{pmatrix}.
\end{eqnarray}
This equation will play a key role in multiple places. The quantity
 $A_{22} - A_{21}A_{11}^{-1} A_{12}\equiv [A /A_{11}]$ will also appear over and over and
 we shall therefore give it its name: it is called the Schur complement $[A /A_{11}]$
 of $A$ with respect to the $11$ block. The block triangular matrices can be trivially inverted and we arrive at a "block $LDU$" decomposition $A=LDU$ in terms of a block lower triangular $L$,
 block diagonal $D$ and block upper triangular matrix $U$,
\begin{eqnarray}
\label{eq:lu2-repeat}
\begin{pmatrix}
A_{11} & A_{12} \\
A_{21} & A_{22}
\end{pmatrix} =
\nonumber \\
\begin{pmatrix}
1 & 0 \\
A_{21} A_{11}^{-1} & 1
\end{pmatrix}
\begin{pmatrix}
A_{11} & 0 \\
0 & [A / A_{11}]
\end{pmatrix}
\begin{pmatrix}
1 & A_{11}^{-1} A_{12} \\
0 & 1
\end{pmatrix}.
\end{eqnarray}
Among the various corollaries of this equation, it provides a close form for the
determinant:
\begin{equation}
\label{eq:Schur}
{\rm det} A =  {\rm det}  [A_{11}]  \ {\rm det} [A / A_{11}]
\end{equation}
The Schur complement has many other nice properties, see \cite{nunezfernandez2025} for a discussion. 
For instance one does not need to take the Schur complement directly with respect to en entire block $A_{11}$, one may do it sub-blocks after sub-blocks and if one does so, the order in which one takes the Schur complements does not matter.

\subsubsection{Cross Interpolation}
The cross interpolation formula approximates $A\approx A_{\mathrm{CI}}$ where $A_{\mathrm{CI}}$ is defined as,
\begin{align}
A_{\mathrm{CI}}  = 
\begin{pmatrix}
A_{11} \\ A_{21} \end{pmatrix}
(A_{11})^{-1}
\begin{pmatrix}
A_{11} & A_{12}
\end{pmatrix}
\end{align}
In other words, the Schur complement is the \emph{error} of the cross interpolation,
\begin{align}  
A = A_{\mathrm{CI}} + 
 \begin{pmatrix}
0 & 0 \\ 0 & [A/A_{11}] 
\end{pmatrix}. 
\end{align}
An important remark is that to construct $A_{\mathrm{CI}}$, one does not need to
know anything about $A_{22}$. Indeed, when a matrix is of (low) rank $\chi$, we only need $\chi$ independent vectors (the first matrix in the definition of
$A_{\mathrm{CI}}$) and $\chi$ rows (which tells us how the other vectors decompose in terms of the independent ones). The cross interpolation formula has two important properties: (i) first it is exact when evaluated on the blocks that have been used to construct it 
($A_{11}$, $A_{12}$ and $A_{12}$) as evident in the above equation. We refer to this as the interpolation property. (ii) Second it is exact if
$A_{11}$ is a $\chi\times\chi$ matrix and $A$ is exactly of rank $\chi$. To prove this second assertion, we construct the sub-matrix of $A$ that contains the $11$ block plus a single extra row $i_0$ and a single extra column $j_0$. Using the Schur complement, we have:
\begin{eqnarray}
\left| {\rm det }
\begin{pmatrix}
A_{11} & A_{1j_0} \\
A_{i_01} & A_{i_0j_0}
\end{pmatrix}
\right| =  | {\rm det} A_{11} | \times |A_{i_0j_0} - A_{i_01} A_{11}^{-1} A_{1j_0}|.
\label{eq:proof_maxvol}
\end{eqnarray} 
(with a slight abuse of notations that mixes indexing with block indexing).
The left hand side is zero by definition of $A$ being of rank $\chi$ 
(it is a $(\chi+1)\times (\chi+1)$ matrix) hence $A_{i_0j_0} - A_{i_01} A_{11}^{-1} A_{1j_0} = 0$, i.e. the cross interpolation is exact.

\subsubsection{Practical cross interpolation}
\begin{figure}
	\centerline{	\includegraphics[scale=0.47]{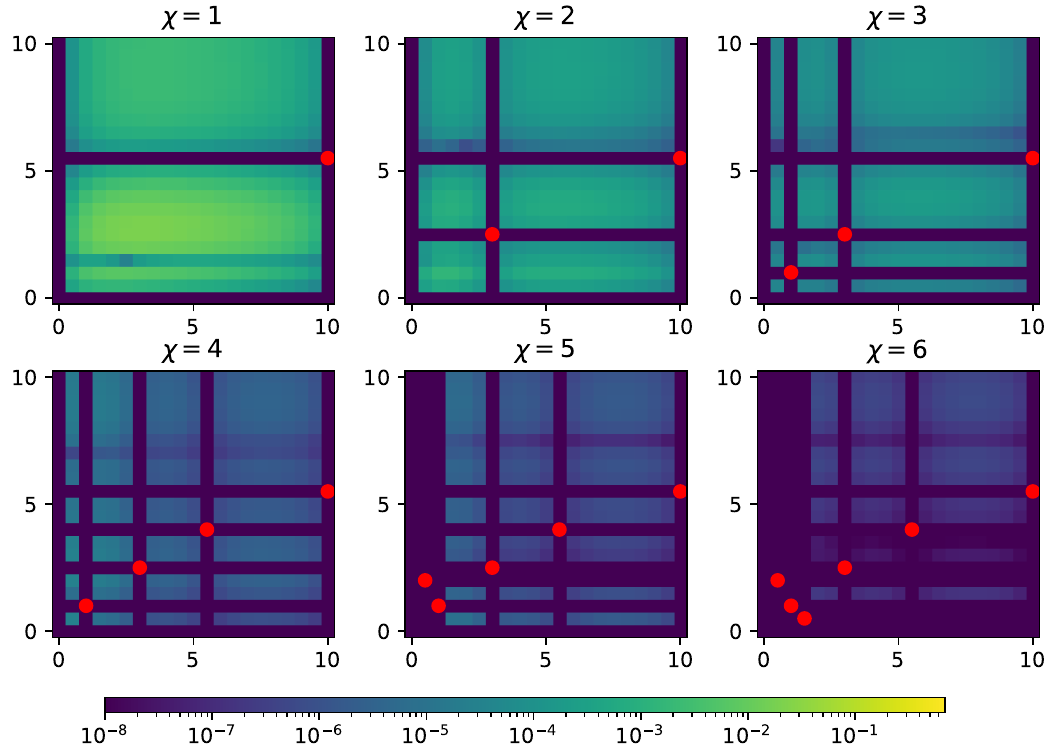}}
	\caption{Error $|A_{ij}-[A_{\mathrm{CI}}]_{ij}|$ versus $i$ and $j$ at different stages of the Cross-Interpolation for a $M\times M$ matrix with $M=20$.
	In this toy example, $A_{ij}=\left(\frac{i/M}{i/M+1}\right)^{4}(1+e^{-(j/M)^{2}})\left[1+(j/M)\cos(j/M)e^{-(j/M)\frac{i/M}{(i/M)+1}}\right]$.
	The red dots indicate the pivots. The $x$ and $y$ axis have been rescaled to be in 
	$[0,10]$. Adapted from Jeannin et al \cite{jeannin2025}.}
	\label{fig:annex_crossi}
\end{figure}

In practice to build up $A_{\mathrm{CI}}$ we need to choose the $A_{11}$ block properly.
Let us introduce the notations that we will use to design the chosen rows and columns.
Let $\mI=\{i_1,i_2, \ldots,i_\chi\}$ (respectively
$\mJ=\{j_1,j_2, \ldots,j_\chi\}$) denote a list of the rows (columns)
of $A$ (that will form the $A_{11}$ block).
Indexing these sets gives the corresponding index: $\mI_a \equiv i_a$ is its a$^{\text{th}}$ element. The list of the indices of
all rows (columns) is denoted $\mathbb{I}=\{1,2, \ldots,M\}$ ($\mathbb{J}=\{1,2, \ldots,N\}$). Following usual programming convention (as in Python/MATLAB/Julia), 
we denote by $A(\mI,\mJ)$ the submatrix
of $A$ comprised of the rows $\mI$ and columns $\mJ$;
$A(\mI,\mJ)_{ab} \equiv A_{\mI_a,\mJ_b}$. We have
\begin{align}
A &= A(\mathbb{I},\mathbb{J}) \\
A_{\mathrm{CI}} &= A(\mathbb{I},\mJ) 
A(\mI,\mJ)^{-1}
A(\mI,\mathbb{J})
\label{eq:matci}
\end{align}
Equation (\ref{eq:matci}) is illustrated graphically in Fig. \ref{fig:ci}. 
The rows and columns of $A(\mI,\mJ)$ are called the
{\it pivots} and $A(\mI,\mJ)$ is the {\it pivot matrix}. 
The pivots are chosen one by one iteratively in such a way as to maximize the
determinant of the matrix $A(\mI,\mJ) = A_{11}$ in order to guarantee that the chosen vectors are truly independent. This is known as the maximum volume (maxvol) principle.
Another way to look at the maxvol principle is that each new pivot is chosen to be the one where the current error of  $A_{\mathrm{CI}}$ is maximum (maxerror) so that adding this pivot brings the largest amount of new information into the approximation. The proof of the equivalence between maxvol and maxerror is in Eq.\eqref{eq:proof_maxvol}. 
A practical example of how the error decreases for the cross extrapolation of a (toy)
matrix is shown in Fig.\ref{fig:annex_crossi}.

The important thing to remember about cross interpolation is that it is given in terms
of slices of the matrix $A$: it is entirely defined in terms of the two lists
$\mI$ and $\mJ$ of the rows and columns of the pivot matrices. 

\subsubsection{Stable evaluation of the cross interpolation}
We need a last ingredient to be able to use cross interpolation in practice. Indeed, as one adds more pivots, the $A_{11}$ matrix becomes increasingly singular so we do not want to
calculate $A_{11}^{-1}$ explicitly as it becomes numerically unstable (even for moderate values of $\chi$). There are several ways to stabilize the evaluation of
\begin{align} 
\begin{pmatrix}
A_{11} \\ A_{21} \end{pmatrix}
(A_{11})^{-1}
\end{align}
The first is to perform a $QR$ factorization of the first matrix, writing
\begin{align} 
\begin{pmatrix}
A_{11} \\ A_{21} \end{pmatrix}
= \begin{pmatrix}
Q_{11} \\ Q_{21} \end{pmatrix} R
\end{align}
The $Q$ matrix being an isometry, it is well conditionned. All the (possibly very small)
singular values of $A_{11}$ are in the triangular matrix $R$ which disappears of the calculation.
Indeed, we have
\begin{align} 
\begin{pmatrix}
A_{11} \\ A_{21} \end{pmatrix}
(A_{11})^{-1} =
\begin{pmatrix}
1 \\ Q_{21}Q_{11}^{-1} \end{pmatrix}
\end{align}

The second way to stabilize this calculation (now our preferred way) is to realize
that Eq.\eqref{eq:lu2-repeat} can be rewritten as
\begin{eqnarray}
\label{eq:prrlu}
\begin{pmatrix}
A_{11} & A_{12} \\
A_{21} & A_{22}
\end{pmatrix} =
\nonumber \\
\begin{pmatrix}
A_{11} & A_{12} \\
A_{21} & A_{21}A_{11}^{-1} A_{12}
\end{pmatrix}
+
\begin{pmatrix}
0 & 0 \\
0 & [A / A_{11}]
\end{pmatrix}
\end{eqnarray}
in other words, if one ignores the Schur complement in Eq.\eqref{eq:lu2-repeat},
one is left with the cross interpolation. We can use Eq.\eqref{eq:lu2-repeat} iteratively,
performing the decomposition pivot after pivot (as stated above, this is legit, the proof can be found in \cite{nunezfernandez2025}) building a decomposition in the form 
$A_{\mathrm{CI}} = LDU$ where $L$ is lower triangular, $D$ diagonal and $U$ upper triangular.
This is nothing but the celebrated $LU$ decomposition used for e.g. inversing matrices.
The only caveat is that it is partial (we stop it after getting the $\chi$ pivots, we don't go all the way through) and it is rank revealing (we use the maxvol criteria to select the pivots). It is the prrLU (partial rank revealing LU) decomposition. But again, this is just 
a neat way to obtain the cross interpolation in a stable way.

\subsection{TCI: Extension of CI to n-dimensional tensors}

We now have everything we need to factorize matrices, we need to extend cross interpolation to tensors. This is what TCI does.

\subsubsection{TCI: naive approach}
We can decompose any tensor using cross interpolation iteratively. We first group together all indices except $\sigma_1$, apply cross interpolation on the resulting matrix and repeat the procedure until the tensor has been entirely factorized. Graphically this (very naive) algorithm has the following form:

\begin{center}
\includegraphics[width=0.45\textwidth]{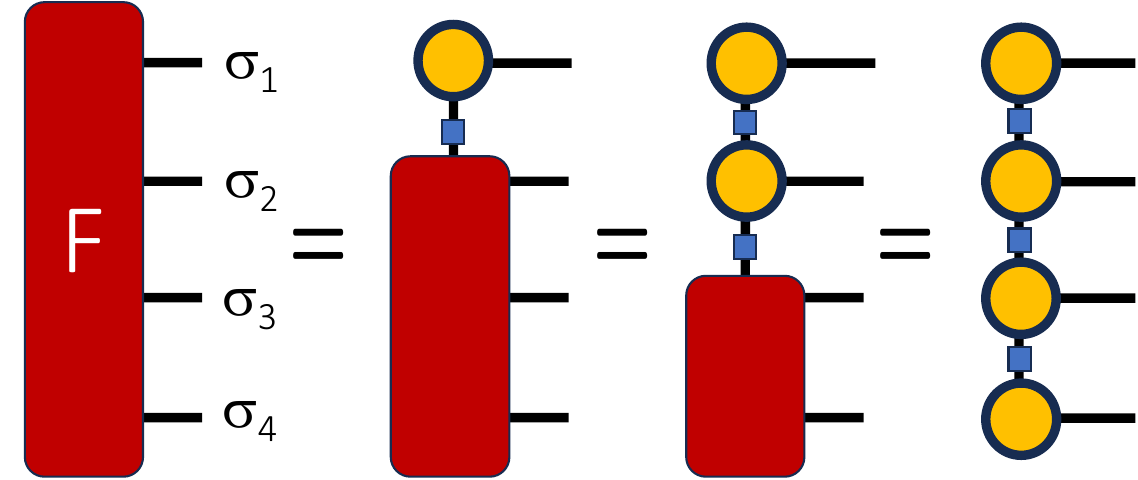}
\end{center}

where the small blue squares stand for the inverse of the pivot matrices. This algorithm is not practical since applying cross interpolation on an exponentially large matrix requires an exponentially large amount of memory and computing time. However, it has
the merit of showing that such a decomposition exists. More interestingly, it shows the structure of the "pivots" of a TCI representation. Indeed, the cross interpolation is defined in terms of the lists $\mI$ and $\mJ$. Now we have one of such list on each side of the pivot matrices (the blue squares above). Each element of this list is now a list itself
that contains the value of the corresponding indices. We call such a list a \emph{multi-index}. More explicitly, we have for our example,

\begin{center}
\includegraphics[width=0.4\textwidth]{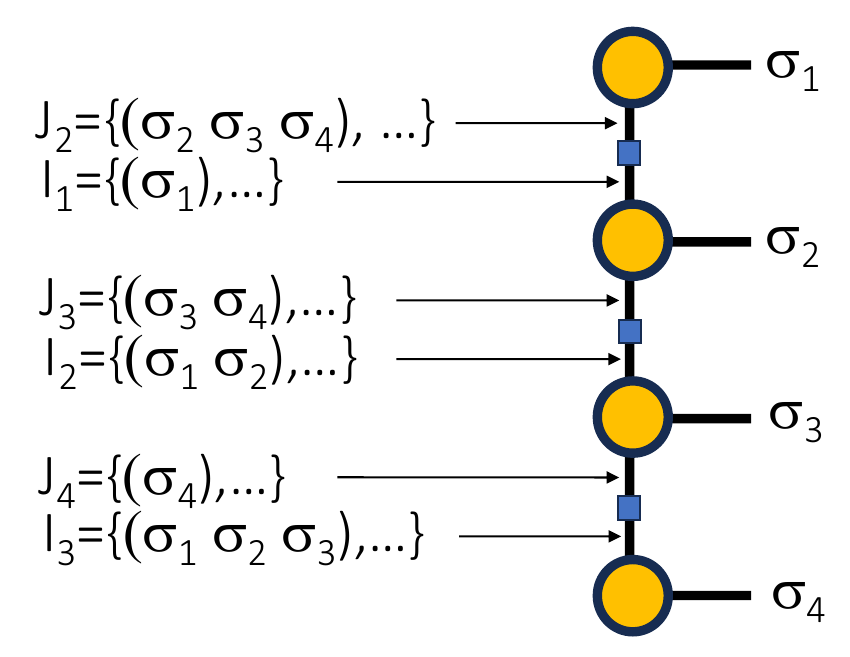}
\end{center}

The most tricky thing about writing a TCI code is to correctly do the book keeping of these lists of lists. 

\subsubsection{TCI: formal form}

Let us introduce our notations a bit more formally. A TCI representation is essentially a MPS but we keep the pivot matrices explicit so that the TCI is entirely made of ``slices'' of the original tensor. For  any $\alpha$ such that $1 \leq \alpha \leq N$,
we consider ``row'' multi-indices $(\sigma_1,\sigma_2, \ldots, \sigma_{\alpha})$ and 
``column'' multi-indices $(\sigma_{\alpha},\sigma_{\alpha +1},
\ldots,\sigma_N)$.
The pivot lists are defined as $\mI_\alpha = \{i_1,i_2, \ldots,i_\chi\}$ 
for the ``rows'' (the multi-indices have size $\alpha$) and  
$\mJ_\alpha = \{j_1,j_2,\ldots,j_\chi\}$ for the ``columns''
(the multi-indices have size $N-\alpha+1$).
For notational convenience, we define $\mI_0$ and $\mJ_{N+1}$ as
singleton sets each comprised of an empty multi-index. Last, we use the symbol $\oplus$ to denote the concatenation of multi-indices: 
\begin{equation} 
  (\sigma_1,\sigma_2, \ldots,\sigma_{\alpha -1}) \oplus (\sigma_\alpha) \oplus (\sigma_{\alpha
  +1}, \ldots,\sigma_n) \equiv (\sigma_1, \ldots,\sigma_N).
\end{equation}

We are now ready to define the TCI representation formally. The definitions are a bit scary looking but they are nothing else than what we obtained above using the naive algorithm.
The blue squares are the pivot matrices $P_\alpha$ defined as,
\begin{equation}
[P_\alpha]_{ij} \equiv F_{[\mI_\alpha]_i \oplus [\mJ_{\alpha+1}]_j}
\end{equation}
(with $P_N=1$ for notation convenience). Likewize the orange three leg tensors $T_\alpha$ are defined as,
\begin{equation}
[T_\alpha]_{i\sigma j} \equiv F_{[\mI_{\alpha-1}]_i
\oplus \sigma \oplus [\mJ_{\alpha+1}]_j}.
\end{equation}
We also introduce the matrix $T_\alpha(\sigma)$ defined as
\begin{equation}
T_\alpha(\sigma)_{ij} \equiv [T_\alpha]_{i\sigma j}
\end{equation}
to make contact with the standard MPS form.
Using these notations, we have,
\begin{equation}
   \label{eq:defTCI}
  F_{\vec{\sigma}} \approx 
  [F_\text{TCI}]_{\vec{\sigma}} \equiv
   \prod_{\alpha=1}^N T_\alpha(\sigma_\alpha) P_\alpha^{-1}.
\end{equation} 
or graphically,

\begin{center}
\includegraphics[width=0.4\textwidth]{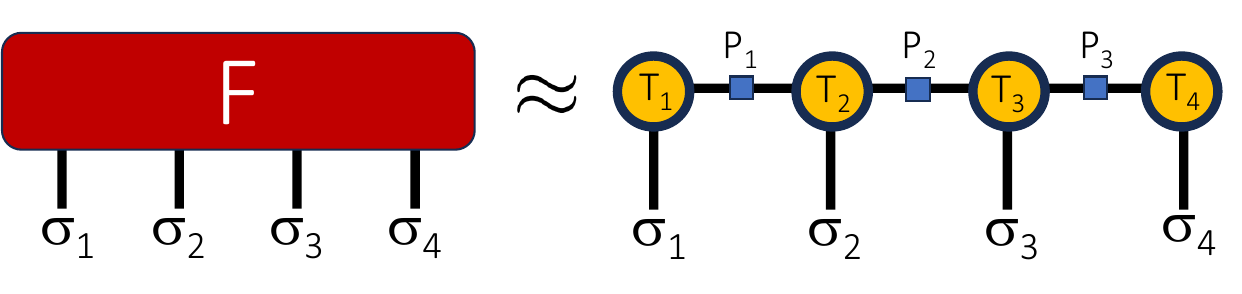}
\end{center}

The TCI representation is defined entirely by the selected sets of
``rows'' and ``columns'' $\mI_\alpha$ and $\mJ_\alpha$, so that
constructing an accurate representation of $F_{\vec{\sigma}}$ amounts to optimizing
the selection of $\mI_\alpha$ and $\mJ_\alpha$ for $1 \leq \alpha \leq
N$. Only $O(N d\chi^2) \ll d^N$ entries of $F_{\vec{\sigma}}$ are used in the
approximation. 

\subsubsection{Practical TCI algorithm} 
\label{sec:TCIalgo}

We start with an initial point $(\sigma_1, \ldots,\sigma_n)$ which we split in $N-1$ different ways $(\sigma_1, \ldots,\sigma_n) = (\sigma_1, \ldots,\sigma_\alpha) \oplus (\sigma_{\alpha+1}, \ldots,\sigma_N)$
 to obtain one element for each of the sets $\mI_\alpha$ and $\mJ_\alpha$.
This yields the initial $\chi=1$ TCI,
which is exact if the tensor $F_{\vec{\sigma}}$ 
factorizes as a product of tensors of one variable. 

To improve on this TCI, we are going to sweep over pairs of tensors 
$(T_\alpha,T_{\alpha+1})$ as is done in two-site DMRG. The sweeping is performed until
convergence. For each pair, we use the following procedure:
First, we introduce yet another tensor, $\Pi_\alpha$ as
\begin{equation}
[\Pi_\alpha]_{i\sigma\sigma' j} \equiv F_{[\mI_{\alpha-1}]_i
\oplus \sigma \oplus \sigma' \oplus [\mJ_{\alpha+2}]_j}.
\end{equation}
Second, we replace $[T_\alpha(\sigma_\alpha) P_\alpha^{-1} T_{\alpha+1}(\sigma_{\alpha+1})]_{ij}$
inside the TCI by $[\Pi_\alpha]_{i\sigma_\alpha\sigma_{\alpha+1} j}$ because the former is a cross interpolation of the latter, hence we might as well use the more precise form.
Next, we continue the cross interpolation of $\Pi_\alpha$ 
(seen as a matrix $[\Pi_\alpha]_{i\otimes\sigma,\sigma' \otimes j}$) 
by adding a new pivot, i.e. one new entry to the list $\mI_\alpha$ and $\mJ_{\alpha+1}$. Graphically, we have:

\begin{center}
\includegraphics[width=0.4\textwidth]{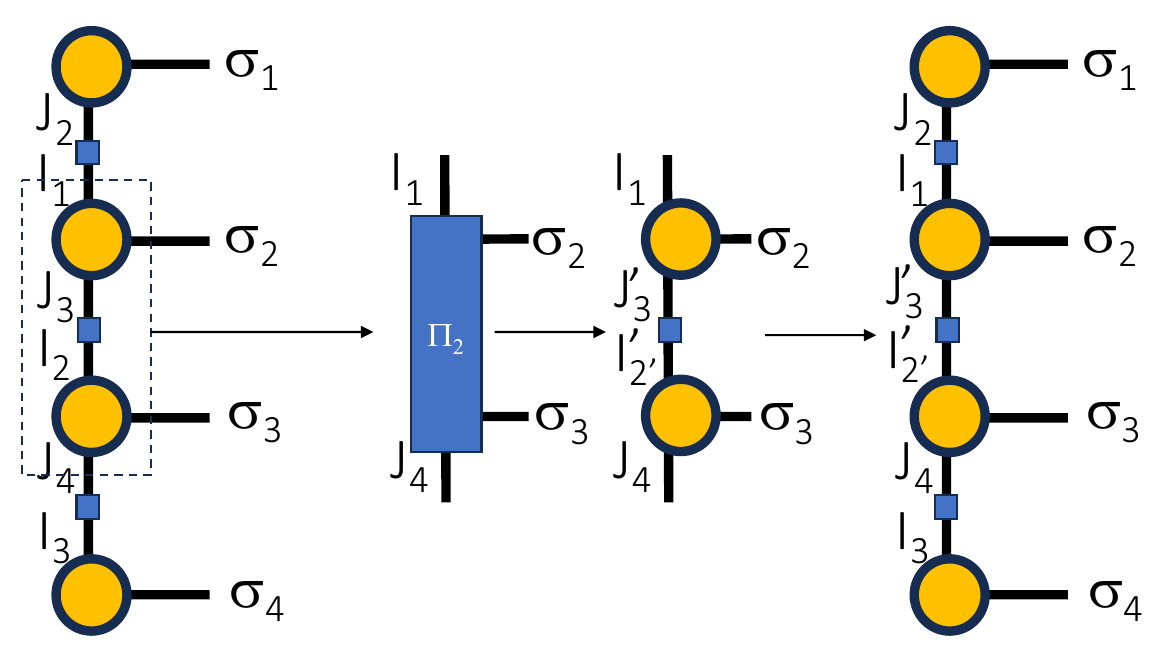}
\end{center}

and that's it, this is a fully functional TCI algorithm (although there are versions that are more suitable for certain purposes).
During the sweeping, we monitor the so-called pivot error between the $\Pi_\alpha$ tensor and its cross interpolation. We stop the iteration when this error is below a certain threshold during an entire sweep.
\begin{equation}
\epsilon_\Pi = \max_{i\sigma\sigma'j} \left|
[\Pi_\alpha]_{i\sigma\sigma' j} - [T_\alpha(\sigma) P_\alpha^{-1} T_{\alpha+1}(\sigma')]_{ij} \right|.
\end{equation}

Now, there is a subtle point that we have swept under the rug: the fact that the error
$\epsilon_\Pi$ is \emph{actually} the error of the TCI approximation for the corresponding pivots,
\begin{align}
\epsilon_\Pi = \max_{i\sigma\sigma'j} \left|
F_{[\mI_{\alpha-1}]_i\oplus\sigma\oplus \sigma' \oplus[\mJ_{\alpha+2}]_j} \right. \nonumber \\ \left.
- [F_\text{TCI}]_{[\mI_{\alpha-1}]_i\oplus\sigma\oplus \sigma' \oplus[\mJ_{\alpha+2}]_j}
 \right|.
\end{align}
Therefore improving the cross interpolation of $\Pi_\alpha$ does indeed improve the TCI approximation itself (at least for these pivots). To prove this point, we need to remember that the cross interpolation is exact on the pivots. We also need to realise that there is a form of ``nesting condition'' that connects the different pivot lists: a pivot $i_\alpha\in\mI_\alpha$ takes the form $i_\alpha = i_{\alpha-1} \oplus \sigma_\alpha$
with $i_{\alpha-1} \in\mI_{\alpha-1}$ (and a similar condition for the $\mJ_\alpha$).
Using these two ingredients, one easily see that there is a telescopic condition for the restriction of the TCI on these pivots. 

Let's see how this works concretely. 
We start by restricting $\sigma_1$ to values that belong to $\mI_1$. For these values,
$T_1$ and $P_1$ cancel due to the interpolation property.
Schematically, it reads,

\begin{center}
\includegraphics[width=0.3\textwidth]{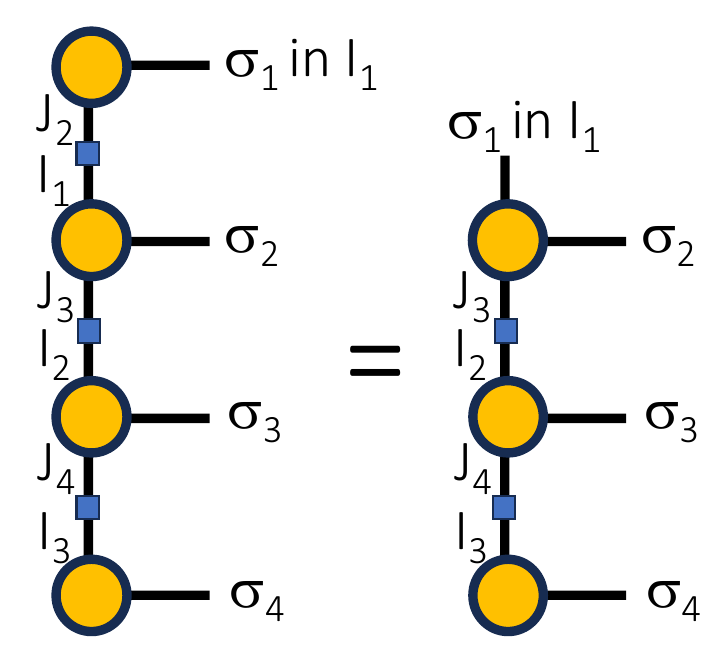}
\end{center}

We continue by requesting that $\sigma_1 \oplus \sigma_2 \in \mI_2$ which we can do
because of the nested condition. The interpolation property implies that,

\begin{center}
\includegraphics[width=0.3\textwidth]{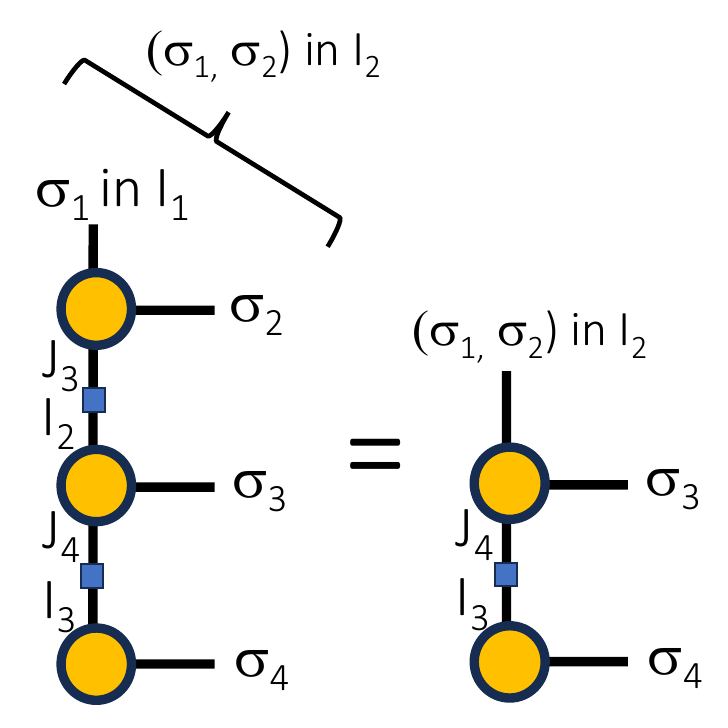}
\end{center}

and we can continue like that down the TCI representation. Since the same thing can be done
with the $\mJ_\alpha$, we can also go up from the bottom of the TCI. See
\cite{nunezfernandez2022} or \cite{nunezfernandez2025} for a more formal proof of the statement.

\subsubsection{Application to integrals}
\label{sec:integrationTCI}
There are many things that one may do with a TCI representation. As stated, it is an entry point to be able to use the many algorithms that have been developped for many-body physics.
For the purpose of these notes, we are interested in multi-dimensional integration. It is an alternative to the Monte-Carlo approach. When it works (the convergence with $\chi$ will depend on the integrand), it compares very favourably to Monte-Carlo in two aspects: the convergence is much faster than what is allowed by the law of large numbers; it is immune to the sign problem that plagues Monte-Carlo whenever the integrand has an oscillatory behaviour. 

In its plainest version, multi-dimensional integration is quite straightforward.
Let us consider a function $f(u_1,...u_n)$ (our integrand). 
We discretize it using a plain quadrature rule with $d$ points per dimension $a_1...a_d$ and the corresponding weight $w_1...w_d$.
For instance, we could use the Gauss-Konrad-21 rule (with $d=21$) or even the trapozeidal rule. We write
\begin{equation}
\int du_1...du_n f(u_1...u_n) \approx \sum_{\sigma_1...\sigma_n} F_{\vec{\sigma}}
w_{\sigma_1}...w_{\sigma_N}
\end{equation}
with
\begin{equation}
F_{\vec{\sigma}} \equiv f(a_{\sigma_1},...,a_{\sigma_n})
\end{equation}
The problem of course is that the sum runs over $d^n$ different configurations which is impractical. This is known as the curse of dimensionality. If, however, we can factorize 
$F_{\vec{\sigma}}$ using TCI, then calculating this sum reduces to $n$ matrix vector multiplication. It becomes essentially trivial,
\begin{equation}
\sum_{\sigma_1...\sigma_n} F_{\vec{\sigma}}
w_{\sigma_1}...w_{\sigma_n} \approx 
 \prod_{\alpha=1}^n \left[\sum_\sigma T_\alpha(\sigma) P_\alpha^{-1}\right].
\end{equation}
This is essentially what we do to calculate our coefficients $Q_n$. There is a small difficulty that we will not discuss in details though: we do not TCI $f(u_1,...u_n)$ (whose rank is very high) but rather $f(u_1,u_2-u_1,u_3-u_2,...u_n-u_{n-1})$ (whose rank is much smaller). The reason is that the later function automatically enforces time-ordering - or more precisely works in a single sector of the ordering (see the corresponding discussion in \cite{nunezfernandez2022}).

\section{Summing up the series (problem C) and results}
\begin{figure} 
   \centerline{ \includegraphics[scale=0.49]{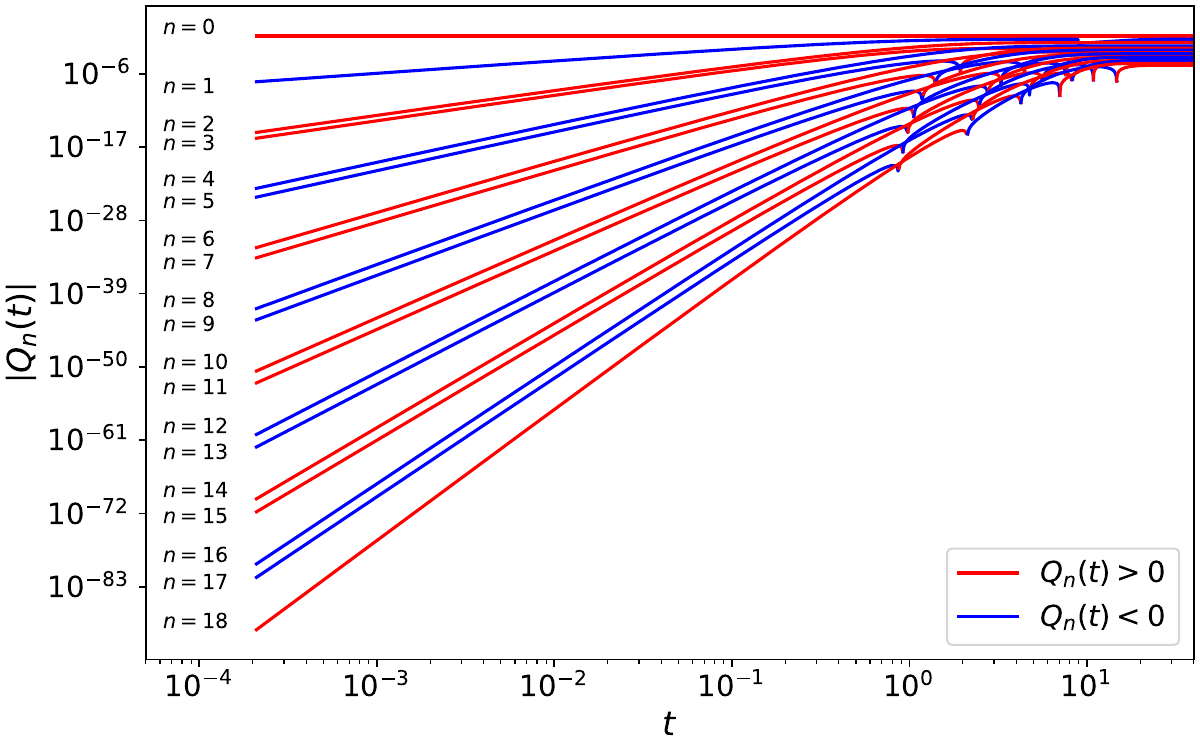} }
   \caption{
      Coefficients of the expansion. Absolute value of the coefficients $Q_n$ as a function of time for the charge ($\epsilon_{d}=0$). Red (blue) portion of curve
corresponds to positive (negative) values of $Q_n$. At large time
$Q_n(t)$ reaches its known exact value (calculated with Bethe ansatz) with high precision. Adapted from \cite{jeannin2025}.
}

\label{fig:coef}
\end{figure}

We are now in position to actually calculate these coefficients $Q_n(t)$. An example of the typical raw data coming out of such a calculation is shown in Fig.\ref{fig:coef} up to order eighteen. Please note the span of the x and y axis: times span four orders of magnitude while the coefficients themselves span almost ninety orders of magnitude. The final result (at infinite time) is correct in this instance with close to eight digits precision. These results are orders of magnitude more precise and faster than what was obtained with previous techniques. We have a last problem to solve (problem C) before we can turn to do a little physics.  

\subsection{Cross extrapolation}
\label{sec:CE}

\begin{figure} 
\includegraphics[scale=0.52]{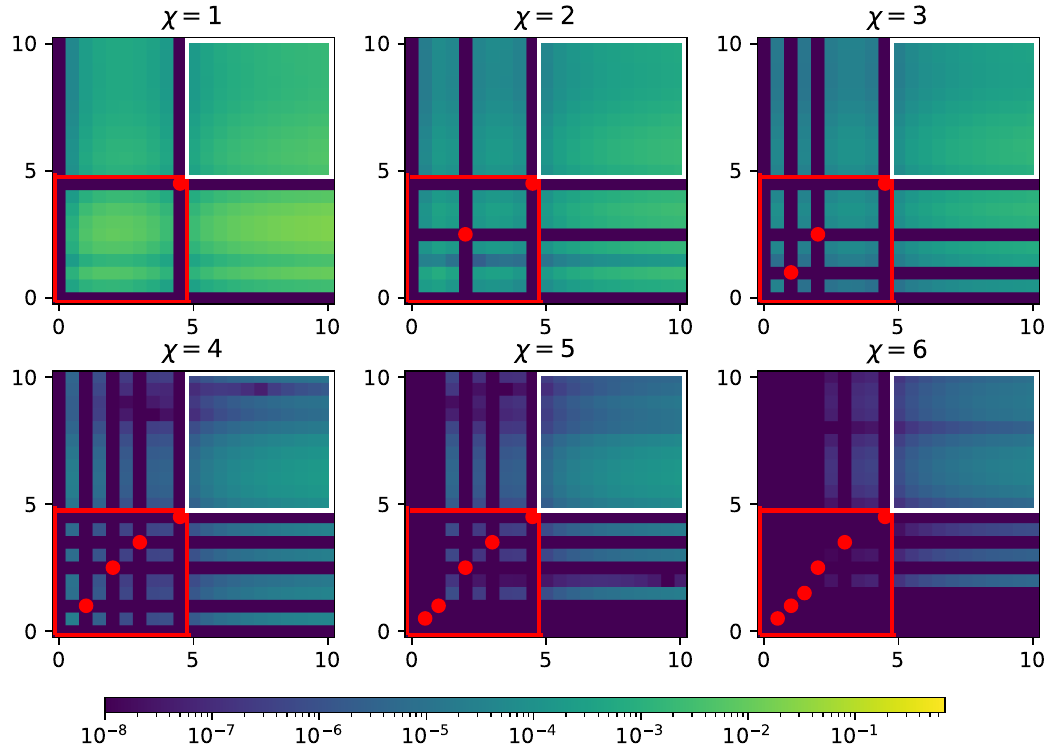} 
\caption{Relative error of the cross-extrapolation versus $i$ and $j$ for a toy matrix
and different number of pivots $\chi=1-6$. The pivots can only be placed inside the red
square in order to extrapolate the results inside the white square. 
Adapted from \cite{jeannin2024}.
\label{fig:ce}}
\end{figure}
Due to the convergence properties of the series of $Q(U,t)$ we can only calculate it
accurately in two regimes: small times (but up to large $U$) and small $U$ 
(but any $t$). We would like to extrapolate our results up to the interesting regime where both $t$ and $U$ are large. Let's first discretize $U$ and $t$ to get a matrix
$Q_{ij} = Q(U_i,t_j)$. Only a few first rows and columns of this matrix are known.

\begin{figure} 
\includegraphics[scale=0.42]{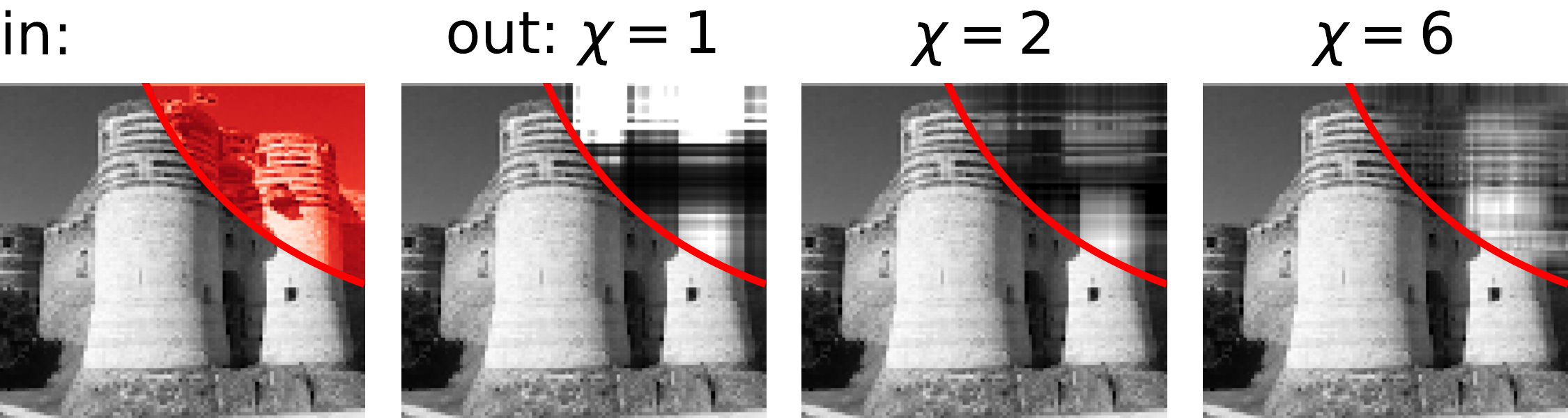} 
\caption{\label{fig:ce2} Example of cross extrapolation for an image (left) whose upper right corner is missing (redish zone). In this instance, it does not work so well. Adapted from \cite{jeannin2024}.}
\end{figure}

The cross-extrapolation algorithm \cite{jeannin2024} is a deceptively simple idea: 
one merely applies the cross-interpolation formula to the matrix $Q_{ij}$ taking advantage of the fact that the formula \emph{does not use} the entire matrix.  
The only difference with what was done in the context of TCI is that one restricts the choices of the pivots to the known sector of the matrix (the $A_{11}$ block).
This is illustrated in Fig.\ref{fig:ce} for a toy example. This is the same matrix as in
Fig.\ref{fig:annex_crossi}, the only difference is that we supposed that the data inside the white square was inaccessible, hence we had to restrict the choice of the pivots to inside the red square. The convergence is not as fast or as stable as in a regular cross interpolation, this is to be expected, but in many instance the method works well and allows one to extrapolate the data further than the initial calculation provides. The error is controled by varying the rank $\chi$ of the extrapolation and the size of the red square region. As in all extrapolation, there is an underlying assumption that must be checked:
the fact that that $Q(U,t)$ is approximately low rank or in other words that for an error level $\epsilon$, there exist a set of $\chi$ one dimensional functions $g_k$ and $h_k$ such that
\begin{equation}
\label{eq:def_low_rank}
| Q(U,t) - \sum_{k=1}^{\chi}g_{k}(U)h_{k}(t) |< \epsilon.
\end{equation}
The approach is very general and could in principle be used in many different situations.
For instance in Fig.\ref{fig:ce2}, we use it to extrapolate an image whose upper right corner is missing.

\subsection{Main results in the stationary limit.}
\begin{figure} 
\centerline{	\includegraphics[scale=0.55]{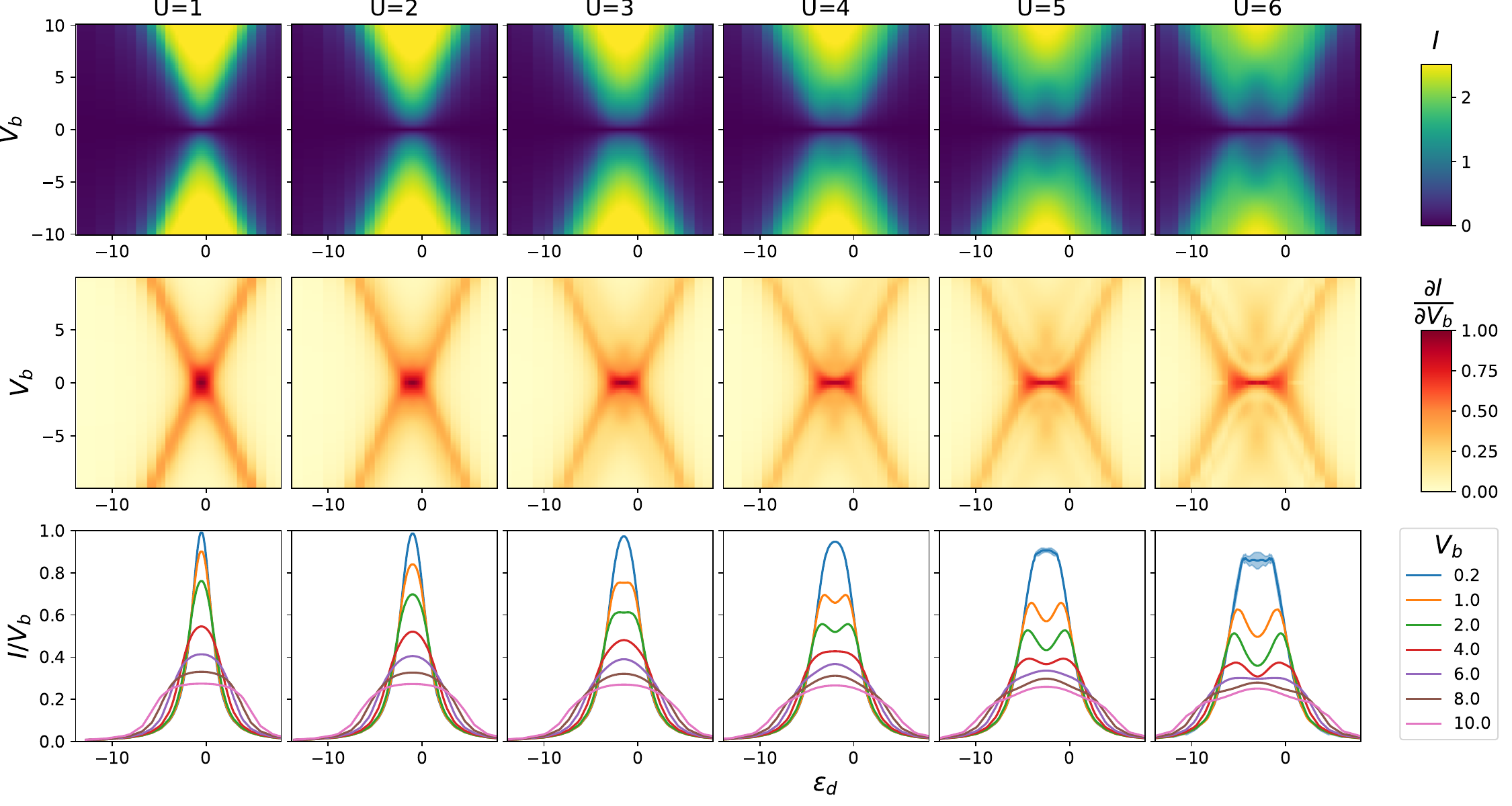}}
\caption{Final result of the calculation after cross extrapolation.
	 Upper panels: color plots of the current versus bias voltage $V_b$ and gate voltage $\epsilon_d$ for SIAM. The diamond like shape shown is known as the "Coulomb diamond". Middle panels: sama data but the differential conductance is shown. the horizontal red lines around $V_b=0$ is known as the "Kondo ridge" and its width is the Kondo energy. Bottom panels: a few horizontal cuts of the same data. Adapted from \cite{jeannin2025}.  \label{fig:diamonds}} 
\end{figure} 

We are now done with the technique itself, it is time to see what it can do in practice and
discuss a little the physics of SIAM. We cannot do justice to SIAM here. The model is very well understood at equilibrium. It is one of the very few correlated model that is both non-trivial (among other things it features the Kondo effect, hence an emerging energy scale) and very well understood (through Bethe ansatz, the numerical renormalisation group, and more).
Calculating its properties out-of-equilibrium though is harder and results in this direction are more recent. On the other hand almost all experiments measure current-voltage
characteristics, hence take place out-of-equilibrium (see the introduction of \cite{jeannin2025} for references). So for a long time we were in a semi-confortable situation where we understood the underlying physics but could not actually calculate the 
observables that were measured. The present technique contributed to bridge this gap.
Fig.\ref{fig:diamonds} shows a snapshot of the results of the calculation featuring both the
``Coulomb diamonds'' (upper panels) and the ``Kondo ridge'' (middle panels), two features that have been observed over and over experimentally. Actually, calculating the Coulomb diamond does not require such a sophisticated technique, it can be perfectly well understood at the semi-classical level. Yet, it is nice to obtain the entire colormap observed experimentally with a single technique and in a controlled way.

\begin{figure} \centerline{
   \includegraphics[scale=0.6]{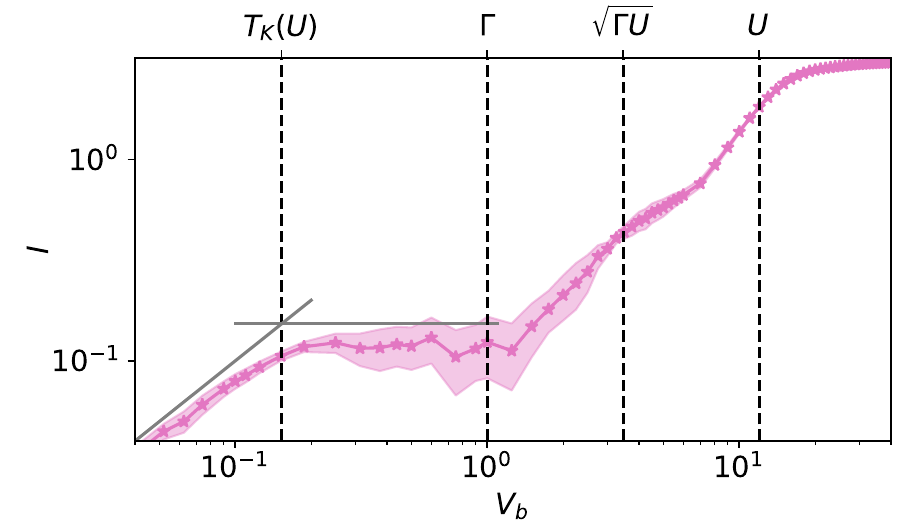}} 
   \caption{ Current $I$ as a function of $V_b$ at $U=12$. The dashed lines
   indicate the position of the energy scales discussed in the text.  The grey
lines correspond to perfect transmission $I=V_b$ and a plateau at $I=T_K(U)$. $N=23$ coefficients were used. The Kondo temperature was extracted from Bethe Ansatz.
Adapted from\cite{jeannin2025}.
}
\label{fig:Kondo_I_Vb_regimes}
\end{figure}

Let's finish by pushing the technique to its (current) limits which is calculate
$I(V_b)$ in the middle of the Kondo ridge at $U=12$. We are going to span three orders of magnitude in bias voltage and cross four different energy scales:
\begin{itemize}
\item The largest one is the charging energy $U$. For $V_b>U$ the current must saturates.
\item The scale $\sqrt{\Gamma U}$ is associated to the fluctuations of the charge. for
energies smaller than this scale, the charge can be considered as ``frozen'' and the problem reduces to its spin sector (i.e. essentially to the so-called Kondo model).
\item The width $\Gamma$ of the non-interacting resonance.
\item The Kondo temperature $T_K$. This scale is what makes the whole problem interesting.
$T_K$ decreases exponentially with $U$. For $V_b<T_K$ one expect a perfect transmission
$I = V_b$ (in units where $e^2/h=1$).  
\end{itemize}
The results are shown in Fig.\ref{fig:Kondo_I_Vb_regimes}. We find the expected regimes
at small and large bias voltages. For $T_K < V_b < \Gamma$, we observe a sort of plateau, that still needs to be confirmed given that the error bars are fairly high in this regime.

\subsection{What's to take away?}
I'd like to end with a few comments as to where this is going. We have seen a technique that consists of several subtechniques (non-interacting Green's functions, Wick determinants,
tensor cross interpolation, series reconstruction) which put together allow to solve a quite
challenging problem. None of these techniques are very difficult once someone has taken the trouble to write a proper open source code that properly isolates the corresponding functionality. This is perhaps my first comment: open source code and proper API are key to being able to assemble complex solutions. We need to be able to program at a level of abstraction that is close to the one we use to think about the problem. My second comment refers to TCI. Finding an underlying structure of the problem to be able to solve it has always be central to how physicists work. What I find amazing with TCI is that this algorithm is able to discover this structure for us. This is generally true of learning algorithms including with deep neural networks and it is a major paradigm shift in the way we think about algorithms. My last comment is about the overall idea of teaching the computer to calculate Feynman diagrams. At the time of this writing it is not clear how far we will be able to go in this direction but I am quite awed by the number of radical speedups and paradigm shifts that we have seen since our first paper on this subject ten years ago. More generally, it seems that the field of computational many-body physics as a whole is moving very quickly these days and I would not be surprised if some of our central problems (say calculating the properties of the 2D Hubbard model) would soon switch from 
``about to be solved'' to ``solved actually''.

\clearpage
\bibliographystyle{correl}
\bibliography{example}

\clearchapter

\end{document}